\documentclass[a4paper,fleqn,usenatbib]{mnras}

\usepackage{newtxtext,newtxmath}
\usepackage[T1]{fontenc}
\usepackage{ae,aecompl}

\usepackage{graphicx}	% Including figure files
\usepackage{amsmath}	% Advanced maths commands
\usepackage{amssymb}	% Extra maths symbols

\usepackage{threeparttable}
\usepackage{lipsum}
\usepackage{mathtools}
\usepackage{cuted}

\title[Magnetized disc-outflow symbiosis]{ Role of magnetically dominated disc-outflow symbiosis on bright hard-state black hole sources: ultra-luminous X-ray sources to quasars}

\author[T. Mondal and B. Mukhopadhyay]{
	Tushar Mondal\thanks{mtushar@iisc.ac.in (TM)} and
	Banibrata Mukhopadhyay\thanks{bm@iisc.ac.in (BM)}
	\\
	Department of Physics, Indian Institute of Science, Bangalore 560012, India
}

\date{Accepted 2020 April 22. Received 2020 April 7; in original form 2019 October 19}

\pubyear{2020}

\begin{document}
\label{firstpage}
\pagerange{\pageref{firstpage}--\pageref{lastpage}}
\maketitle

\begin{abstract}
We present optically thin solutions for magnetized, advective disc-outflow symbiosis around black holes (BHs). The main objective is to explain the bright, hard state observations of accreting systems with stellar-mass to supermassive BHs. We include the effects of magnetic fields and radiation counterpart in entropy gradient based on the first law of thermodynamics to represent energy advection. The cooling process includes bremsstrahlung, synchrotron radiation, and inverse Comptonization process. One of our main ventures is to explain some long-standing issues of ultra-luminous X-ray sources (ULXs). The existing physical scenarios to explain their unusual high luminosity are either the existence of the missing class of intermediate-mass BH (IMBH), or super-Eddington accretion around a stellar-mass BH. However, most ULXs with steep power-law spectrum can be well explained through super-Eddington accretion, while the existence of IMBH is indeed disputed extensively. Nevertheless, the interpretation of ULXs with a hard power-law dominated state remains mysterious. Here we show that our magnetically dominated disc-outflow symbiosis around rapidly spinning stellar-mass BHs can achieve such large luminosity even for sub-Eddington accretion rate. The magnetic field at the outer zone of the advective flow is more than the corresponding Eddington limit. Such field becomes dynamically dominant near the BH through continuous accretion process due to flux freezing, but maintaining its Eddington limit. This unique field configurations enhance the synchrotron, and synchrotron self-Comptonization process to achieve very large luminosity. Through the same mechanism, our solutions for supermassive BHs can explain the unusual large luminosity of ultra-luminous quasars. 
\end{abstract}


%%%%%%%%%%%%%%%%%%%%%%%%%%%%%%%%%%%%%%%%%%%%%%%%%%

%%%%%%%%%%%%%%%%% BODY OF PAPER %%%%%%%%%%%%%%%%%%

\section{Introduction} \label{sec:introduction}

Galactic black hole candidates (BHCs) are known to pass through different X-ray spectral states based on their X-ray luminosity and spectral shape. Typically, the most familiar states are high/soft (HS), low/hard (LH), and very high/steep power-law (VH/SPL) states \citep{2004MNRAS.355.1105F, 2006ARA&A..44...49R}. In HS state, the X-ray spectrum is dominated by multicolor disc blackbody component with characteristic temperature $\sim 1$ keV. The disc luminosity $L$ is quite below the Eddington luminosity $L_{\text{Edd}}$ $(L \sim 0.1 \ L_{\text{Edd}})$, which is defined as
\begin{equation}
L_{\text{Edd}}=\frac{4\pi cGM_{BH}}{\kappa_{es}}\simeq 1.4\times 10^{38} \left(\frac{M_{BH}}{M_{\odot}} \right) \ \text{erg s}^{-1}, \label{eq:L_Edd}
\end{equation}
where $M_{BH}$ is the mass of the black hole (BH), $G$ the Newton's gravitation constant, $c$ the speed of light, and $\kappa_{es}$ the electron scattering opacity. The X-ray spectrum in LH state is dominated by a hard power-law component with an exponential high-energy cutoff at $\sim 200$ keV. The luminosity in this state is comparatively low $(L<0.01 \ L_{\text{Edd}})$ and the photon index of the power-law component is $\Gamma \sim 1.4-1.8$. In VH/SPL state, the blackbody component and power-law component become comparable. This state is observed at higher luminosity $(L\sim 1-10\ L_{\text{Edd}})$ and photon index is steeper $(\Gamma \sim 2.5)$ than that found in LH state.

It is believed that these different spectral states correspond to the different accretion geometries depending on mass accretion rates $(\dot{m})$ and, hence, matter density. When the accretion rate is large enough ($\dot{m}\sim 0.1$, in Eddington unit), a standard geometrically thin, optically thick, radiatively efficient \citep{1973A&A....24..337S} Keplerian disc is formed. The HS state is governed by this accretion paradigm. For larger accretion rate ($\dot{m}\gtrsim 1$), a slim disc \citep{1988ApJ...332..646A} is formed and the corresponding spectral state is VH/SPL. In this regime, photon trapping may take place due to very large optical thickness and, hence, some disc luminosity could be advected into the BH. On the other side, when the accretion rate is very small ($\dot{m}\lesssim 0.01$), the flow becomes optically thin. The Coulomb coupling between ions and electrons becomes very weak. The generated heat in the dissipation process partly can be stored as entropy rather than being radiated out. The different classes of such radiatively inefficient accretion flows could be described by many models: an ion-supported torus \citep{1982Natur.295...17R} or an advection-dominated accretion flow \citep[ADAF;][]{1995ApJ...452..710N} or a two-component accretion flow \citep[TCAF;][]{1995ApJ...455..623C} or an adiabatic inflow-outflow solutions \citep[ADIOS;][]{1999MNRAS.303L...1B} or a convection-dominated accretion flow \citep[CDAF;][]{2000ApJ...539..798N, 2000ApJ...539..809Q}  or a general advective accretion flow \citep[GAAF;][]{2010MNRAS.402..961R}. All these models can explain the hard spectral states.

Apart from these most familiar canonical states, there are large number of BH sources which are very bright but their spectra are hard power-law dominated. The RXTE
%The Rossi X-Ray Timing Explorer ($RXTE$) 
observations of the BHC GX 339-4 show the luminosity up to $\sim 0.3 \ L_{\text{Edd}}$ in its hard spectral state \citep{2008PASJ...60..637M}. Also a significant fraction of ultra-luminous X-ray sources (ULXs) in their hard spectral states are observed with luminosity in the range of $3\times 10^{39}-3\times 10^{40} \ \text{erg s}^{-1}$. The true nature of such sources (Antennae X-11, X-16, X-42, X-44, \citealt{2009ApJ...696.1712F}; NGC 1365, \citealt{2009ApJ...695.1614S}; M99 X1, \citealt{2006MNRAS.372.1531S}; M82 X42.3+59, \citealt{2010ApJ...710L.137F}; Holmberg IX, \citealt{2009ApJ...702.1679K}) remains mysterious over decades. Not only stellar mass BHs, a large fraction of supermassive BHs also appears very bright in their hard power-law dominated states. Some of these such sources are ultra-luminous quasars (e.g. PKS 0743-67, \citealt{2005ApJ...633L..89P}; HS 1946+7658, \citealt{1992A&A...253L...5H}), luminous BL Lac objects (e.g. PKS 0301-243, 1ES 0502+675, \citealt{2011ApJ...743..171A}).

Exactly in this line, our proposal comes in. Our group already initiated to explain the importance of strong magnetic fields in BH accretion sources in the advective paradigm \citep{2015ApJ...807...43M, 2018MNRAS.476.2396M, 2019MNRAS.482L..24M,2019MNRAS.486.3465M}. In \cite{2019MNRAS.482L..24M}, for the first time in literature, we suggest a plausible mechanism to explain the unusually large luminosity for certain ULXs. We showed that hard-state ULXs are magnetically powered sub-Eddington, advective accretors around stellar-mass BH. However, we did not include the cooling processes explicitly therein. In this paper, we address a magnetized disc-outflow symbiosis with explicit cooling to explain the unusual large luminosity of those astrophysical BH sources in their hard-spectral states, where large scale strong magnetic fields are assumed to play the underlying roles. Magnetic fields may influence the accretion dynamics, remove angular momentum from in-falling matter, enhance cooling mechanism through synchrotron and synchrotron self-Comptonization (SSC) process, help in the formation of outflows/jets, and play the key role in collimation, as well as to stabilize the powerful jets. Note that the origin of angular momentum transport in accretion disc had been a puzzle over decades until \citet{1991ApJ...376..214B} pointed out the role of the magnetorotational instability (MRI) within differentially rotating disc. This powerful shearing instability mediated by a weak magnetic field can maintain magnetic turbulence. The Maxwell stress generated by turbulent magnetic fields transports angular momentum efficiently in ionized accretion flows. However, apart from small-scale magnetic fields, accretion disc can carry large-scale magnetic fields as well. The transport within the plunging region of a BH occurs due to large-scale magnetic torques \citep{1999ApJ...522L..57G, 2000ApJ...528..161A, 2006ApJ...651.1023R}. Even outside the plunging region, transport may happen via large-scale magnetic fields \citep{1974Ap&SS..28...45B, 2003PASJ...55L..69N, 2018MNRAS.476.2396M}. 

The origin, as well as the strength of large-scale strong magnetic fields in BH accretion flows, is still not well understood. However, a strong correlation between hard spectral states, powerful jets, and dynamically dominant magnetic fields has been found. Recent observations confirm the signatures of dynamically dominant magnetic fields in the vicinity of BHs \citep{2013Natur.501..391E, 2014Natur.510..126Z}. The magnetically dominated accretion flows have been studied in different versions of models, namely magnetically arrested disc \citep[MAD;][]{2003PASJ...55L..69N}, or magnetically choked accretion flow \citep{2012MNRAS.423.3083M}. Such models are based on the idea originally proposed by \cite{1974Ap&SS..28...45B}, where strong vertical poloidal fields are dragged towards the central BH by continuous accretion process. This has also been verified numerically via pseudo-Newtonian magnetohydrodynamics simulations \citep{2003ApJ...592.1042I} and via general relativistic magnetohydrodynamics (GRMHD) simulations \citep{2011MNRAS.418L..79T, 2012MNRAS.423.3083M}. Apart from MAD, alternate magnetically supported accretion flow models have been discussed, in which the magnetic field geometry is dominated by radial and toroidal fields both, and it operates only when the accretion rate is relatively high \citep{2006PASJ...58..193M, 2010ApJ...712..639O, 2016MNRAS.459.4397S}.
 
In this light of the discussion, we propose a general advective, sub-Eddington disc-outflow symbiotic model in the presence of large-scale strong magnetic fields. Unlike MAD, the advection of both toroidal and poloidal magnetic fields is happening here. This model can explain the bright, hard spectral state of BH sources of mass ranging from stellar-mass to supermassive scales. Most importantly, we address the hidden nature of hard-state ULXs without incorporating the existence of intermediate-mass BHs.

The paper is organized as follows. In the next section, we model the coupled general advective disc-outflow symbiosis, including thermodynamic properties in the presence of magnetic fields and radiation. The solution procedure, along with appropriate boundary conditions, is mentioned in Section~\ref{sec:method}. In Section~\ref{sec:result}, we discuss our results, which cover the disc flow behaviours, magnetic field properties, and the energetics of this magnetized accretion flows. Various observational implementations of our results are given in Section~\ref{sec:observations}. In Section~\ref{sec:discussion}, we discuss the possible origin and the strength of magnetic fields in this advective paradigm. We also discuss the implication of our model, particularly for very luminous hard-state BH sources. Finally we end with conclusion in Section~\ref{sec:conclusion}.

%In 1972, Bekenstein proposed an idealized engine, namely Geroch-Bekenstein engine, that makes use of the extreme gravitational potential of a BH (BH) to convert mass to energy with almost perfect efficiency. The practical realization of this engine is very difficult in astrophysical systems. Generally astrophysical BHs do convert mass to energy via accretion process with modest efficiencies. Later the idea came to address the importance of large scale dipole magnetic field in an accretion flow. Following this idea, it was suggested and also verified numerically that this efficient conversion is practically possible in the presence of large scale poloidal magnetic field. Such an efficient accretion phenomenon is named as Magnetically Arrested Disk (MAD).

\section{MODELING THE COUPLED MAGNETIZED DISC-OUTFLOW SYSTEM} \label{sec:model}

\subsection{General equations of magnetized advective accretion flow}

We formulate a general magnetized disc-outflow symbiotic model around BHs in geometrically thick, optically thin, advective accretion framework. Unlike previous exploration \citep[e.g.][]{2004ApJ...616..669K}, we consider all possible viscous stresses as well as large-scale magnetic stress. Indeed, it was initiated earlier by us, without considering the cooling effect explicitly \citep{2019MNRAS.482L..24M}. As a consequence, we could not comment on the luminosity of the system explicitly. In this disc-outflow symbiosis, we adopt the cylindrical coordinate system assuming a steady and axisymmetric flow such that $\partial / \partial t \equiv \partial / \partial \phi \equiv 0$. This 2.5-dimensional quasi-spherical advective flow describes the inner part of the accretion where gravitational force dominates over the centrifugal force of the flow, unlike standard Keplerian disc. All the dynamical flow parameters, namely, radial velocity $(v_{r})$, specific angular momentum $(\lambda)$, outflow or vertical velocity $(v_{z})$, adiabatic sound speed $(c_{s})$, fluid pressure $(p)$, mass density $(\rho)$, radial $(B_{r})$, azimuthal $(B_{\phi})$, and vertical $(B_{z})$ components of magnetic field, are functions of both radial and vertical coordinates. In this formalism, we use pseudo-Newtonian potential, given by \citet{2002ApJ...581..427M}, which mimics certain features of general relativity quite accurately. The other key features in this model are as follows. First, the vertical flow is included here explicitly, and unlike \citet{2010ApJ...713..105B}, it is coupled to the other flow parameters through fundamental equations of motion. The outflows are more likely to emanate from the hot puffed up region of the accretion flow. Second, we include the effect of viscosity by taking care of all possible components of viscous shearing stress. Third, the effect of large-scale magnetic field geometries is included explicitly, unlike  \citet{2005A&A...434..839M} or \citet{2010MNRAS.402..961R}. Fourth, we include the effects of magnetic fields and radiation counterpart in the computation of entropy gradient and adiabatic exponents based on the first law of thermodynamics. 

Throughout our calculations, we express all the flow variables in dimensionless units. The radial and vertical coordinates are expressed in units of the gravitational radius $r_{g}=GM_{BH}/c^{2}$. Any flow velocities are expressed in units of $c$, the specific angular momentum in $GM_{BH}/c$, the fluid pressure, mass density, and magnetic fields, accordingly, to make all the variables dimensionless. Hence, the continuity equation and the components for momentum balance equation are respectively,
\begin{equation}
\frac{1}{r} \frac{\partial}{\partial r}\left(r\rho v_{r}\right)+\frac{\partial}{\partial z}\left(\rho v_{z}\right)=0, \label{continuity}
\end{equation}
\begin{multline}
v_{r}\frac{\partial v_{r}}{\partial r}+v_{z}\frac{\partial v_{r}}{\partial z}-\frac{\lambda^{2}}{r^{3}}+\frac{1}{\rho}\frac{\partial p}{\partial r}+F=\frac{1}{\rho}\frac{\partial W_{rz}}{\partial z}\\
+\frac{1}{4 \pi  \rho}\left[-\frac{B_{\phi}}{r}\frac{\partial}{\partial r}\left(rB_{\phi}\right)+B_{z}\left(\frac{\partial B_{r}}{\partial z}-\frac{\partial B_{z}}{\partial r}\right)\right], \label{rad_momentum}
\end{multline}
\begin{multline}
v_{r}\frac{\partial \lambda}{\partial r}+v_{z}\frac{\partial \lambda}{\partial z}=\frac{r}{\rho}\left[\frac{1}{r^{2}}\frac{\partial}{\partial r}\left(r^{2}W_{r\phi}\right)+\frac{\partial W_{\phi z}}{\partial z}\right]\\
+\frac{r}{4\pi \rho}\left[\frac{B_{r}}{r}\frac{\partial}{\partial r}\left(rB_{\phi}\right)+B_{z}\frac{\partial B_{\phi}}{\partial z}\right],
\end{multline}
\begin{multline}
v_{r}\frac{\partial v_{z}}{\partial r}+v_{z}\frac{\partial v_{z}}{\partial z}+\frac{1}{\rho}\frac{\partial p}{\partial z}+\frac{Fz}{r}=\frac{1}{r\rho}\frac{\partial}{\partial r}\left(rW_{rz}\right)\\
+\frac{1}{4 \pi  \rho}\left[B_{r}\left(\frac{\partial B_{z}}{\partial r}-\frac{\partial B_{r}}{\partial z}\right)-B_{\phi}\frac{\partial B_{\phi}}{\partial z}\right]. \label{vertical_momentum}
\end{multline}
Here $F$ is the magnitude of the force corresponding to the gravitational potential for a rotating BH in the pseudo-Newtonian framework \citep{2002ApJ...581..427M}, as given by
\begin{equation}
F=\frac{\left( r^{2}-2a\sqrt{r}+a^{2} \right)^{2}}{r^{3}\left[ \sqrt{r}(r-2)+a \right]^{2}},
\end{equation}
where $a$ is the Kerr-parameter. $W_{ij}$s are the components for viscous shearing stress tensor. Following standard practice, the $W_{r\phi}$ component is written using standard-disc \citep{1973A&A....24..337S} prescription with proper modification due to advection \citep{1996ApJ...464..664C} as $W_{r\phi} = \alpha (p+\rho v_{r}^{2})$, where $\alpha$ prescribes the turbulent viscosity. The other components can be simplified in terms of $W_{r\phi}$ as $W_{\phi z} \simeq \frac{z}{r}W_{r\phi}$ and $W_{rz} \simeq \frac{z}{r} \alpha W_{r\phi}$ \citep{2009RAA.....9..157G}.

The magnetohydrodynamics (MHD) flow provides two fundamental equations for magnetism, namely, the equation for no magnetic monopole and the induction equation. These are respectively, 
\begin{equation}
\frac{1}{r}\frac{\partial}{\partial r}\left(rB_{r}\right)+\frac{\partial B_{z}}{\partial z}=0,
\end{equation}

\begin{equation}
\frac{\partial}{\partial z}\left[r\left(v_{z}B_{r}-v_{r}B_{z}\right)\right]=0,
\end{equation}
\begin{equation}
\frac{\partial}{\partial r}\left(v_{r}B_{\phi}-\frac{\lambda B_{r}}{r}\right)=\frac{\partial}{\partial z}\left(\frac{\lambda B_{z}}{r}-v_{z}B_{\phi}\right),
\end{equation}
\begin{equation}
\frac{\partial}{\partial r}\left[r\left(v_{z}B_{r}-v_{r}B_{z}\right)\right]=0.
\end{equation}
Here, the induction equation is written in the limit of very large magnetic Reynolds number, which is the case for an accretion disc.

\subsection{Thermodynamics of the gas: in the presence of radiation and magnetic field}

The first law of thermodynamics allows us to calculate the entropy gradient in terms of temperature and density gradients. This entropy gradient changes in different types of accretion processes depending on the detailed balance of heating, cooling, and advection. Different accretion rate, and hence matter density, provides the information of cooling mechanisms. Also magnetic field plays an important role in heating, as well as, cooling processes. Hence, the equation of state for a mixture of perfect gas and radiation in the presence of magnetic field is
\begin{equation}
p_{t}=p+p_{m}=p_{g}+p_{r}+p_{m}=\frac{\rho k_{B}T}{\mu m_{p}}+\frac{1}{3}aT^{4}+\frac{B^{2}}{8\pi},
\end{equation}
%p_{t}=p+p_{m}=p_{g}+p_{r}+p_{m}=\frac{\rho k_{B}T}{\mu m_{p}}+\frac{1}{3}aT^{4}+\frac{B^{2}}{8\pi},
where $p_{t}$ is the total pressure, $k_{B}$ the Boltzmann constant, $\mu$ the mean molecular weight, $m_{p}$ the proton mass and $a$ the Stefan constant. The three different pressures: gas $(p_{g})$, radiation $(p_{r})$ and magnetic $(p_{m})$ can be written in terms of two fundamental parameters $\beta$ and plasma-$\beta$ $(\beta_{m})$ given by 
\begin{equation}
p_{g}=\beta p \ , \ p_{r}=(1-\beta)p \ , \ p_{m}=(\beta /\beta_{m})p .
\end{equation}
Throughout we take the parameter $\beta = (p_{g}/p)$ to be independent of $r$, unlike the parameter $\beta_{m}=(p_{g}/p_{m})$. Note that we keep the same definition of $\beta$ as discussed in the context of gas-radiation mixture \citep[e.g.,][]{1983psen.book.....C}.

The internal energy per unit mass of the system, in the presence of magnetic field, is
\begin{equation}
U=\frac{3}{2}\frac{\rho k_{B}T}{\mu m_{p}}V+aT^{4}V+\frac{B^{2}}{4\pi}V,
\end{equation}
where $V$ is the volume of unit mass of gas, and hence $\rho V=1$. Using first law of thermodynamics and flux-freezing assumption, the entropy gradient is then given by
\begin{equation}
Tds=\frac{p}{\rho}\left[ \left( 12-\frac{21}{2}\beta \right) \frac{dT}{T}-\left(4-3\beta+\frac{1}{3}\frac{\beta}{\beta_{m}} \right) \frac{d\rho}{\rho} \right]. \label{entropy}
\end{equation}
Following \cite{1967aits.book.....C}, we can define the adiabatic exponents $\Gamma_{1}$ and $\Gamma_{3}$ by the equations 
\begin{equation}
\frac{dp}{p}+\Gamma_{1}\frac{dV}{V}=0 \  \ \text{and} \ \frac{dT}{T}+(\Gamma_{3}-1)\frac{dV}{V}=0.
\end{equation}
For such gas-radiation mixture in the presence of magnetic fields, these exponents are
\begin{equation*}
\Gamma_{1}=\frac{32-24\beta -3\beta^{2}+2\beta (4-3\beta)/(3\beta_{m})}{24-21\beta}, 
\end{equation*}
\begin{equation}
\Gamma_{3}=\frac{32-27\beta +2\beta /(3\beta_{m})}{24-21\beta}. \label{exponents}
\end{equation}
In the absence of magnetic fields $(\beta_{m}\rightarrow \infty)$, the entropy gradient and $\Gamma_{1}$ and $\Gamma_{3}$ will reduce to their respective hydrodynamical forms as given by \citet{1988ApJ...332..646A} and \citet{1983psen.book.....C} respectively.

\subsection{Radiation mechanisms for two-temperature plasma}

For simplicity, we assume the gas consists of ions and electrons. From charge neutrality, the number densities of ions and electrons are equal. However, the plasma in this advective paradigm behaves like a two-temperature system due to the large mass difference between ions and electrons. Hence we allow the electron temperature $T_{e}$ and ion temperature $T_{i}$ to be different and the gas pressure of the accreting gas can be written as 
\begin{equation}
p_{g}=\beta p = \frac{\rho k_{B}T_{i}}{\mu_{i} m_{p}}+\frac{\rho k_{B}T_{e}}{\mu_{e} m_{p}},
\end{equation}
where $\mu_{i}$ and $\mu_{e}$ are the effective molecular weights for ions and electrons respectively. Since ions are much heavier than electrons, we normally expect all the generated heats due to viscous and/or magnetic dissipations primarily act on ions. Some part of this heat may transfer from ions to electrons via Coulomb coupling. Finally electrons take part in radiating heat. The separate energy equations for ions and electrons are considered by taking detailed balance of heating, cooling and advection. The energy equation for ions is given by
\begin{multline}
\frac{24-21\beta}{2(4-3\beta)} \left[ v_{r}\left\lbrace\frac{\partial p}{\partial r}-\Gamma_{1}\frac{p}{\rho}\frac{\partial \rho}{\partial r}\right\rbrace+v_{z}\left\lbrace\frac{\partial p}{\partial z}-\Gamma_{1}\frac{p}{\rho}\frac{\partial \rho}{\partial z}\right\rbrace \right] \\
=Q^{+}-Q^{ie}.
\label{eq:energy1}
\end{multline}
Here, the radial and vertical advection of the ion flows are represented by the first and second terms of the left-hand side of the equation~(\ref{eq:energy1}). The right-hand side of equation~(\ref{eq:energy1}) represents the difference between the rate of energy gained (generated via dissipation process, $Q^{+}$) and lost (transfer from ions to electrons, $Q^{ie}$) by ions per unit volume. $Q^{+}$ consists both the viscous and magnetic dissipation parts as $Q^{+}=Q^{+}_{\text{vis}}+Q^{+}_{\text{mag}}$. The rate of generated heat per unit volume due to viscous dissipation is given by 
\begin{equation}
Q^{+}_{\text{vis}}=\alpha \left(p+\rho v_{r}^{2}\right)\frac{1}{r}\left[\frac{\partial \lambda}{\partial r}-\frac{2\lambda}{r}+\frac{z}{r}\frac{\partial \lambda}{\partial z}+\alpha z\left(\frac{\partial v_{r}}{\partial z}+\frac{\partial v_{z}}{\partial r}\right)\right].
\end{equation}
On the other hand, the magnetic heating is basically due to the abundant supply of magnetic energy and/or due to the annihilation of the magnetic fields. The rate of generated heat per unit volume due to magnetic dissipation is given by \citep{2019MNRAS.482L..24M}
\begin{multline}
Q^{+}_{\text{mag}}=\frac{1}{4\pi}\left[B_{r}B_{z}\left(\frac{\partial v_{r}}{\partial z}+\frac{\partial v_{z}}{\partial r}\right)+B_{\phi}B_{r}\left(-\frac{1}{r}\frac{\partial \lambda}{\partial r}+\frac{2\lambda}{r^{2}}\right)\right .\\ \left . +\frac{B_{\phi}B_{z}}{r}\frac{\partial \lambda}{\partial z}\right].
\end{multline}
The energy transfer from ions to electrons happens through Coulomb coupling. This term behaves like heating term in the energy balance equation for electrons. The volume transfer rate of energy from ions to electrons is given by \citep{2000ApJ...529..978B}
\begin{multline}
q^{ie}=\frac{8\sqrt{2\pi}e^{4}n_{i}n_{e}}{m_{i}m_{e}}\left(\frac{T_{e}}{m_{e}}+\frac{T_{i}}{m_{i}} \right)^{-3/2} \ln (\Lambda)(T_{i}-T_{e}) \\ \text{ergs\ cm$^{-3}$\ s$^{-1}$},
\end{multline}
where $\ln (\Lambda)\approx 20$ is the Coulomb logarithm. Here the dimensionless form $(Q^{ie})$ is linked through $q^{ie}=Q^{ie}c^{11}/(G^{4}M_{BH}^{3})$. The energy equation for electrons is then given by
\begin{multline}
\frac{24-21\beta}{2(4-3\beta)} \left[ v_{r}\left\lbrace\frac{\partial p_{e}}{\partial r}-\Gamma_{1}\frac{p_{e}}{\rho}\frac{\partial \rho}{\partial r}\right\rbrace+v_{z}\left\lbrace\frac{\partial p_{e}}{\partial z}-\Gamma_{1}\frac{p_{e}}{\rho}\frac{\partial \rho}{\partial z}\right\rbrace \right] \\
=Q^{ie}-Q^{-}, 
\label{eq:energy2}
\end{multline}
where $Q^{-}$ represents the radiative cooling rate through electrons via different cooling processes including bremsstrahlung $(q_{br}^{-})$, synchrotron $(q_{syn}^{-})$, and inverse Comptonization processes of bremsstrahlung radiation $(q_{br,C}^{-})$, as well as synchrotron soft photons $(q_{syn,C}^{-})$. The radiative cooling rate per unit volume is
\begin{equation*}
q^{-}=Q^{-}c^{11}/(G^{4}M_{BH}^{3})=q_{br}^{-}+q_{syn}^{-}+q_{br,C}^{-}+q_{syn,C}^{-}.
\end{equation*}
Various cooling formalisms are adopted from \cite{1995ApJ...452..710N} and \cite{2010MNRAS.402..961R} and can be written as
\begin{multline}
q_{br}^{-}=1.4\times 10^{-27}n_{i}n_{e}T_{e}^{1/2}(1+4.4\times 10^{-10}T_{e}) \\ \text{ergs\ cm$^{-3}$\ s$^{-1}$},
\end{multline}
\begin{equation}
q_{syn}^{-}=\frac{2\pi}{3c^{2}}kT_{e}\frac{\nu _{c}^{3}}{R} \ \text{ergs\ cm$^{-3}$\ s$^{-1}$},
\end{equation}
\begin{multline}
q_{br,C}^{-}=3\eta_{1}q_{br}^{-}\left\{ \left(\frac{1}{3}-\frac{x_{c}}{3\theta_{e}}\right)-\frac{1}{\eta_{2}+1}\left[\left(\frac{1}{3}\right)^{\eta_{2}+1}- \left(\frac{x_{c}}{3\theta_{e}}\right)^{\eta_{2}+1} \right] \right\} \\ \text{ergs\ cm$^{-3}$\ s$^{-1}$},
\end{multline}
\begin{equation}
q_{syn,C}^{-}=q_{syn}^{-}\eta_{1}\left[1-\left(\frac{x_{c}}{3\theta_{e}}\right)^{\eta_{2}} \right] \ \text{ergs\ cm$^{-3}$\ s$^{-1}$},
\end{equation}
where $\nu_{c}$ is the synchrotron self-absorption cut-off frequency. This can be determined from the relation 
\begin{equation}
\nu_{c}=\frac{3}{2}\nu_{0}\theta_{e}^{2}x_{m},\ \text{with}\ \nu_{0}= \frac{eB}{2\pi m_{e}c},\ \text{and}\ \theta_{e}=\frac{k_{B}T_{e}}{m_{e}c^{2}}.
\label{eq:cut_off_frequency}
\end{equation}
The parameter $x_{m}$ is computed numerically at every radius $R \ (=rGM_{BH}/c^{2})$ following \cite{1995ApJ...452..710N}. The Comptonization energy enhancement factor $(\eta)$ gives the average change in energy of a photon between injection and escape. The factor $(\eta -1)$ times the mean flux of escaping photons provides an extra cooling due to Comptonization. \cite{1991ApJ...369..410D} have given an approximate prescription for this $\eta$, as 
\begin{multline}
\eta =1+\frac{P(A-1)}{(1-PA)}\left[1-\left(\frac{x}{3\theta_{e}}\right)^{-1-\ln P/\ln A} \right] \\
\equiv 1+\eta_{1}\left[1-\left(\frac{x}{3\theta_{e}}\right)^{\eta_{2}} \right],
\end{multline}
where
\begin{equation*}
x=h\nu /m_{e}c^{2}, \ P=1-\exp(-\tau_{es}), \ \text{and} \ A=1+4\theta_{e}+16\theta_{e}^{2}.
\end{equation*}
Here, $x$ is the photon energy at injection, $P$ the probability that an escaping photon is scattered, $\tau_{es}$ the scattering optical depth, and $A$ the mean amplification factor in the energy of a scattered photon.

\section{SOLUTION PROCEDURE} \label{sec:method}

In this disc-outflow coupled region, we make a reasonable assumption that within the inflow regime, the vertical variation of any dynamical flow parameters (say, $Y$) is much less than that with radial variation, which allows us to choose $\partial Y/\partial z\approx sY/z$, where the constant $s$ is scale parameter for that corresponding variable. The magnitude of $s$ is very small compared to unity $(|s|\sim 0.01)$.
Note that the BH accretion is transonic. The sub-sonic matter far away from the BH passes through sonic/critical point and becomes super-sonic. Following standard practice \citep[e.g.][]{2010MNRAS.402..961R}, we combine all the above equations to express $dv_{r}/dr$ in terms of all dynamical variables and independent variable $r$, as 
\begin{equation}
\frac{dv_{r}}{dr}=\frac{\mathcal{N}}{\mathcal{D}}.
\end{equation} 
At `critical radius' $r=r_{c}$, $\mathcal{D}$ becomes zero. To capture smooth solutions around such a point,
$\mathcal{N}$ must be vanished therein. Any variables with subscript `$c$' refers to the values of that respective variables at the critical radius. 
Here, we prescribe
\begin{multline}
B_{rc}=\sqrt{4\pi \rho_{c}}\frac{c_{sc}}{f_{r}\sqrt{3}},\ B_{\phi c}=k_{\phi}B_{rc},\ B_{zc}=k_{z}B_{rc}, \\ \text{and} \ v_{zc}=\frac{c_{sc}}{f_{1}},\label{parameter}
\end{multline} 
with $c_{sc}=\sqrt{\frac{\Gamma_{1} p_{c}}{\rho_{c}}}$, and $\sqrt{3}$ as normalization factor. The relative strength between different  magnetic field components at the critical radius is determined by the constants $f_{r}$, $k_{\phi}$, and $k_{z}$, whereas, the vertical/outflow velocity is controlled by  $f_{1}$.

The matter density and the outflow velocity at the critical point are prescribed through the following information. Integrating the continuity equation, we obtain the total mass accretion rate $\dot{M}$, as given by
\begin{equation}
\int_{0}^{r}\int_{0}^{2\pi}\int_{-h}^{h}\left[ \frac{\partial}{\partial r}\left(r\rho v_{r}\right)+\frac{\partial}{\partial z}\left(r\rho v_{z}\right) \right] \ dr d\phi dz = \dot{M}. \label{acc_rate}
\end{equation}
The first term on the left-hand side of the equation~(\ref{acc_rate}) signifies the rate at which the radial mass flux changes, whereas the second term indicates the vertical mass flux rate. Hence this total mass accretion rate can decouple into inflow rate $\dot{M_{a}}$ and outflow rate $\dot{M_{j}}$, which is used at critical point to prescribe the information of vertical velocity and mass density at that location in this disc-outflow symbiotic model, as $\dot{M}=\dot{M_{a}}+\dot{M_{j}}$. Therefore, the inflow rate and outflow rate can be read as
\begin{equation}
\dot{M_{a}}=\int_{0}^{h} 4\pi r\rho v_{r}\ dz \  \ \text{and} \ \dot{M_{j}}=\int 4\pi r\rho v_{z}\ dr +c_{j} ,
\end{equation}
where the constant $c_{j}$ is determined from appropriate boundary condition. Throughout in our computation and discussion, we express the mass accretion rate $(\dot{m})$ in units of the Eddington rate: $\dot{M}_{\text{Edd}}\equiv L_{\text{Edd}}/(\eta c^{2})=1.39\times 10^{18} (M/M_{\odot})$ g s$^{-1}$, where $\eta$ $(\approx 0.1$ here) is the radiative efficiency factor, i.e., $\dot{m}=\dot{M}/\dot{M}_{\text{Edd}}$.  

Now there is an upper limit to the amount of magnetic field that a disc around a BH can sustain. Also, there is an upper bound to the amount of outflow velocity at the critical point to obtain physical solutions. These constraints are fixed through the parameters $f_{r}$, $k_{\phi}$, $k_{z}$, and $f_{1}$, as mentioned in equation (\ref{parameter}). Below we prescribe the procedure to fix these parameters using certain physical conditions at critical point.
\begin{figure*}
	\center
	\includegraphics[width=17cm]{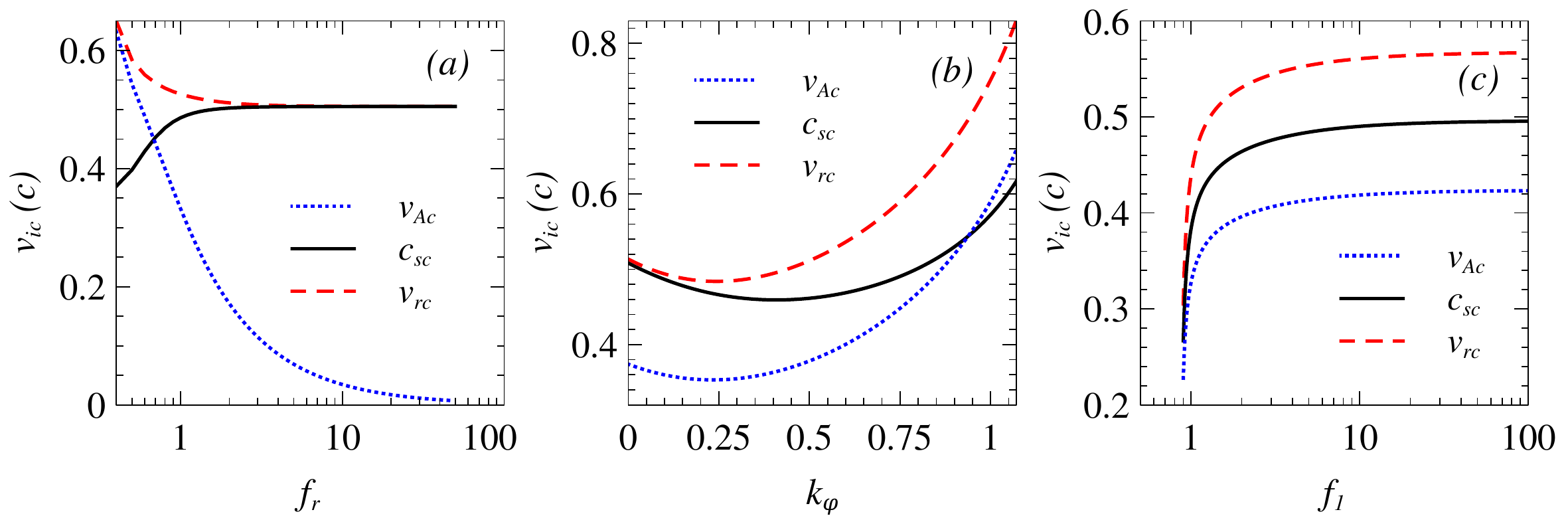}\caption{Variations of radial velocity, sound speed, and Alfv$\acute{\text{e}}$n velocity at critical point as functions of  $(a)$ inverse of the magnetic field strength, $(b)$ ratio of azimuthal to radial magnetic field, and $(c)$ inverse ratio of vertical velocity to sound speed. The other model parameters are $M_{\text{BH}}=20 M_{\odot}$, $\dot{m}=0.05$, and $\alpha = 0.01$. 
	} 
	\label{fig:parameters}
\end{figure*}  
The upper limit of the magnetic field strength is constrained through the parameter $f_{r}$, as shown in Figure~\ref{fig:parameters}$(a)$. Basically, different velocity components, say,  $v_{rc}$, $c_{sc}$, and Alfv$\acute{\text{e}}$n velocity $(v_{Ac})$, at the critical point have been computed through $\mathcal{N}=0$ and $\mathcal{D}=0$. To check the dependence on $f_{r}$, first we have to fix other parameters $k_{\phi},$ $k_{z},$ and $f_{1}$. Then we fix physically reasonable $f_{r}$, and check the dependence on the arbitrariness of the other parameters. Figure \ref{fig:parameters}$(a)$ infers that the influence of magnetic field is negligible at a large value of $f_{r}$ in which the system behaves like a simple hydrodynamic one with Mach number $(v_{rc}/c_{sc})$ at critical point tends to become unity. With decreasing $f_{r}$, the influence of magnetic field starts to play an important role. The Alfv$\acute{\text{e}}$n velocity overcomes the medium sound speed below $f_{r}=0.7$. We have not obtained any physical solutions in this regime and, hence, we have fixed $f_{r}>0.7$ for our computation. By keeping $f_{r}=0.8$, we look for the dependence on $k_{\phi}$, as described in Figure \ref{fig:parameters}$(b)$. Again for the same reason, we have not obtained any physical solutions with $k_{\phi}\gtrsim 0.9$. Now the dependence on $f_{1}$ has been given in Figure \ref{fig:parameters}$(c)$ using $f_{r}=0.8$, and $k_{\phi}= 0.6$. Smaller the value of $f_{1}$, larger the outflow velocity at the critical point is. The physical parameters become imaginary and, hence, unphysical for $f_{1}<0.9$. This way one can obtain these parameters to choose maximum allowable magnetic field components and outflow velocity at the critical location. Note that various $s_{i}$ are chosen accordingly for a self-consistent solution. Any wrong choice of $s_{i}$ produces unphysical measure of a variable. 

Now at $r=r_{c}$, $dv_{r}/dr=0/0$. Applying l'Hospital rule and some algebraic simplification, we find that the velocity gradient at the critical point has two values. These two values define the nature of the critical point \citep{1990ttaf.book.....C, 2018MNRAS.476.2396M}. When both are complex, the critical point is `spiral'-type. When both are real, and of same sign, the critical point is `nodal'-type. When both are real, and of opposite sign, the critical point is `saddle'-type. The negative slope at `saddle'-type critical point indicates the accretion solution, whereas the positive one refers to wind solution. Due to two-temperature prescription in this advective paradigm, we need to adequately specify the electron temperature $T_{ec}$ and specific angular momentum $\lambda_{c}$ at $r_{c}$. One can solve the eigenvalue problem for $\lambda_{c}$ at the critical point $r_{c}$ using a relaxation method self-consistently through physical boundary conditions \citep[e.g.,][]{1997ApJ...476...49N}. However, this method is little bit challenging for the presently considered complex system with explicit cooling, magnetic field profiles, and vertical variation of flow variables. Hence, we choose an alternate but equivalent approach, in which a set of $\lambda_{c}$ and $r_{c}$ is prescribed, but operating within a certain range depending on the spin of the BH, and then solve the equations. The maximum allowed values of $\lambda_{c}$ for different spin parameters of a BH are already mentioned in the literature \citep[e.g.,][]{2003ApJ...586.1268M, 2015ApJ...807...43M}. We need to adjust these parameters along with the relative dependence of magnetic field geometries, to capture the full dynamical solutions connecting the outer boundary to the BH event horizon through $r_{c}$. The simultaneous solution for $\mathcal{N}=0$, $\mathcal{D}=0$ and the equation for vertical momentum balance (equation~\ref{vertical_momentum}) provide self-consistently the numerical values for $z_{c}$, $c_{sc}$ and $v_{rc}$ at $r=r_{c}$. The outer boundary corresponds to the radius $r=r_{out}$, where $\lambda=\lambda_{K}$. $\lambda_{K}$ is the Keplerian angular momentum per unit mass of the flow where the centrifugal force balances the gravity. The inner boundary corresponds to the event horizon of the BH, at which radius the velocity becomes the speed of light (in our unit unity). We also have to supply $M_{BH}$, $\dot{M}$, $\alpha$, and $\beta$. 

\section{Coupled disc-outflow physics} \label{sec:result}

\subsection{Disc dynamics: stellar mass black holes}
\begin{figure*}
	\center
	\includegraphics[width=16 cm]{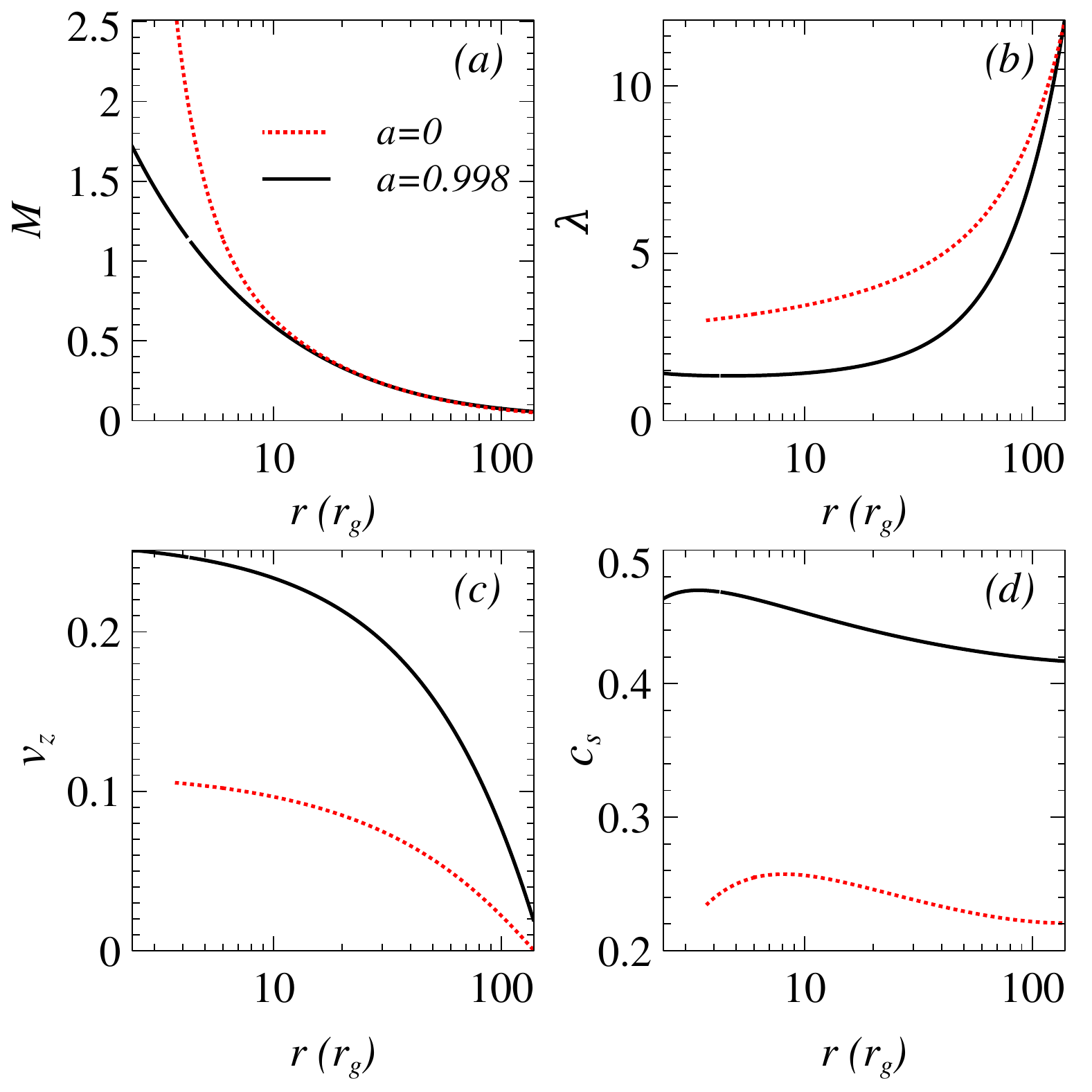}\caption{Variations of $(a)$ Mach number, $(b)$ specific angular momentum, $(c)$ vertical/outflow velocity, and $(d)$ sound speed, as functions of radial coordinate. The model parameters are $M_{\text{BH}}=20 M_{\odot}$, $\dot{m}=0.05$, and, $\alpha = 0.01$. 
	} 
	\label{fig:dynamics}
\end{figure*}
We address here how the large-scale strong magnetic fields can influence the transport of angular momentum, as well as, outflow dynamics in the advective paradigm. The disc is here quasi-spherical and hot puffed-up. In Fig.~\ref{fig:dynamics}, we show how the fundamental flow parameters vary throughout this disc-outflow coupled system. The location of the critical point and specific angular momentum value at such critical location are respectively $r_{c}=6.0$, $\lambda_{c}=3.184$ for non-rotating BH $(a=0)$, and $r_{c}=4.2$, $\lambda_{c}=1.3385$ for fast-rotating BH $(a=0.998)$. The typical values for vertical scale parameter $`s$' for the corresponding flow variables $v_{r},\ \lambda,\ v_{z},\ B_{r},\ B_{\phi},\ B_{z},\ p,$ and $\rho$ are respectively $s_{1}=-0.014,\ s_{2}=-0.01,\ s_{3}=0.04,\ s_{4}=-0.03842,\ s_{5}=-0.03,\ s_{6}=0.01,\ s_{7}=-0.035,$ and $s_{8}=-0.03$. The magnetic field configuration at the critical location is $B_{\phi c}=1.8 B_{rc}$, $B_{z c}=-B_{rc}/5$ when $a=0$, and $B_{\phi c}=0.6 B_{rc}$, $B_{z c}=-B_{rc}/5$ when $a=0.998$.

Fig.~\ref{fig:dynamics}$(a)$ describes the Mach number $(M)$ profile, which is defined as the ratio of the radial velocity to the sound speed. It indicates that very far away from the BH, the matter is sub-sonic and is independent of BH's spin. As it advances towards the central BH, it becomes super-sonic near the event horizon of a BH. The sonic locations for non-spinning and fast-spinning BHs are respectively $r=6.6269$ and $r=5.049$. The inner boundary of our solution is defined as the location where effective matter velocity $\left(v=\sqrt{v_{r}^{2}+(\lambda /r)^{2}+v_{z}^{2}}\right)$ reaches the speed of light. This location does not overlap exactly with the corresponding location of the event horizon of a BH, because of the use of pseudo-Newtonian potential instead of full GR computation. Fig.~\ref{fig:dynamics}$(b)$ describes the outward transport of the angular momentum in this flow. Very far away from the BH, the transition radius between the Keplerian and sub-Keplerian flows is the outer boundary for our solutions. The Keplerian angular momentum basically signifies the profile in which the centrifugal force balances the gravitation force of the BH. As matter falls towards the central BH, it loses angular momentum to form the disc. However, it is very difficult to model self-consistently the transition region where $\lambda/\lambda_{K}=1$. This is because the set of equations used to model a hot advective flow is not strictly continued to be valid to explain cold, optically thick, Keplerian disc model. Note that, in this paper, we do not intend to address this transition zone, rather we concentrate on the hot advective part. In Fig.~\ref{fig:dynamics}$(c)$, we show the profile for the vertical/outflow velocity. At outer zone, it is almost negligible as usual. The outflow is basically emanated from the hot puffed-up region of the advective disc-outflow surface. Our model is valid vertically up to the upper surface of the disc-outflow coupled region, above which flow will decouple and accelerate to form plausible jets. The vertical velocity near vicinity of the BH increases from $0.1053$ when $a=0$ to $0.2512$ when $a=0.998$. The corresponding sound speed profile is shown in Fig.~\ref{fig:dynamics}$(d)$. The medium sound speed at around horizon increases from $0.2352$ when $a=0$ to $0.4637$ when $a=0.998$. The increase of $c_{s}$ with spin indicates that the temperature of the system is higher for fast-spinning BHs. All such changes suggest the possibility that the energetics of such advective flows may strongly depend on the BH spin. 

\begin{figure*}
	\center
	\includegraphics[width=16 cm]{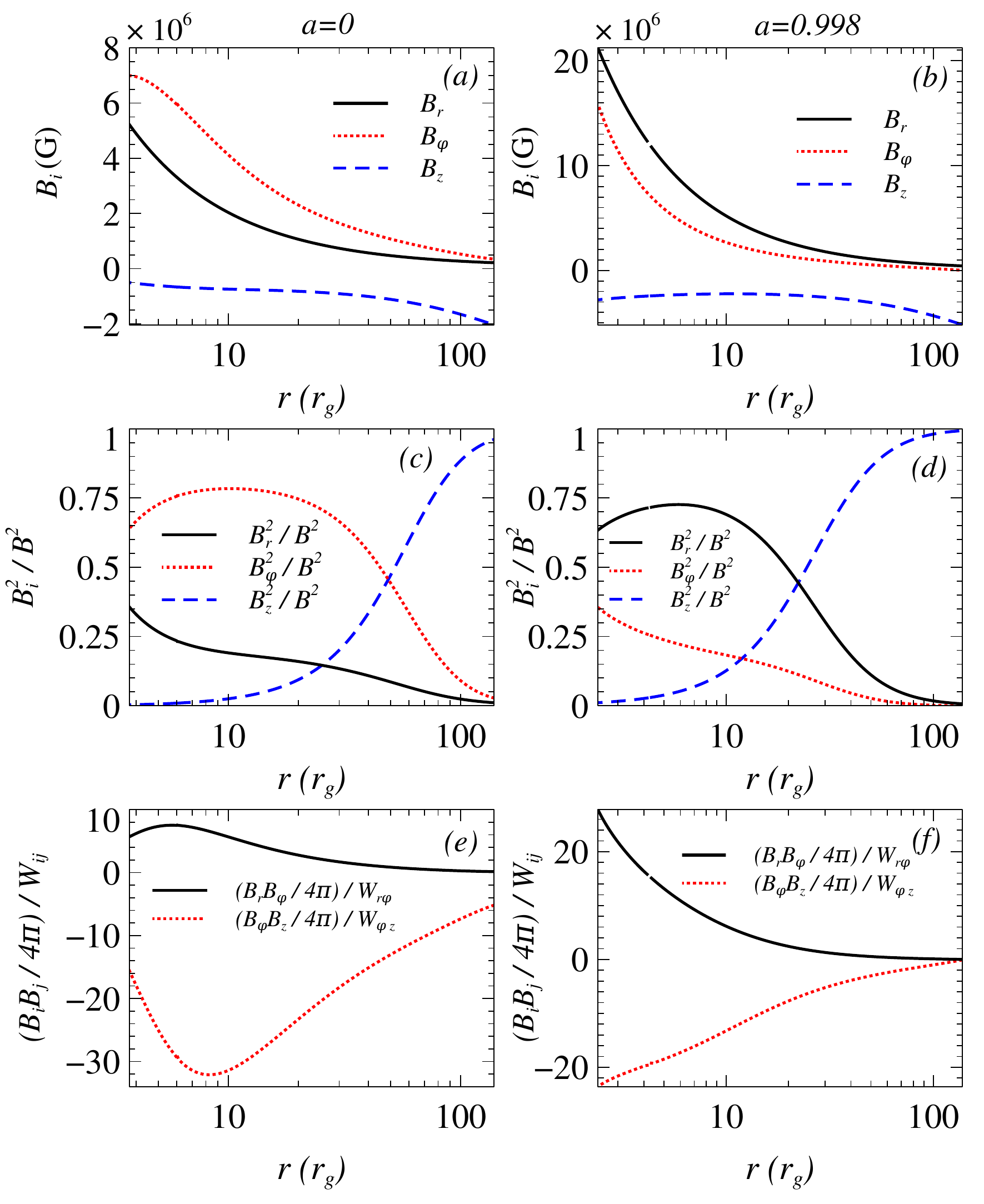}\caption{Variations of $(a)$ magnetic field components, $(c)$ relative field strength, $(e)$ the ratios of magnetic to viscous stresses, as functions of radial coordinate, for $a=0$. $(b)$, $(d)$, and $(f)$ depict the same as $(a)$, $(c)$, and $(e)$ respectively, except for $a=0.998$. The model parameters are same as in Fig~\ref{fig:dynamics}. 
	} 
	\label{fig:field}
\end{figure*}
\begin{figure}
	\center
	\includegraphics[width=\columnwidth]{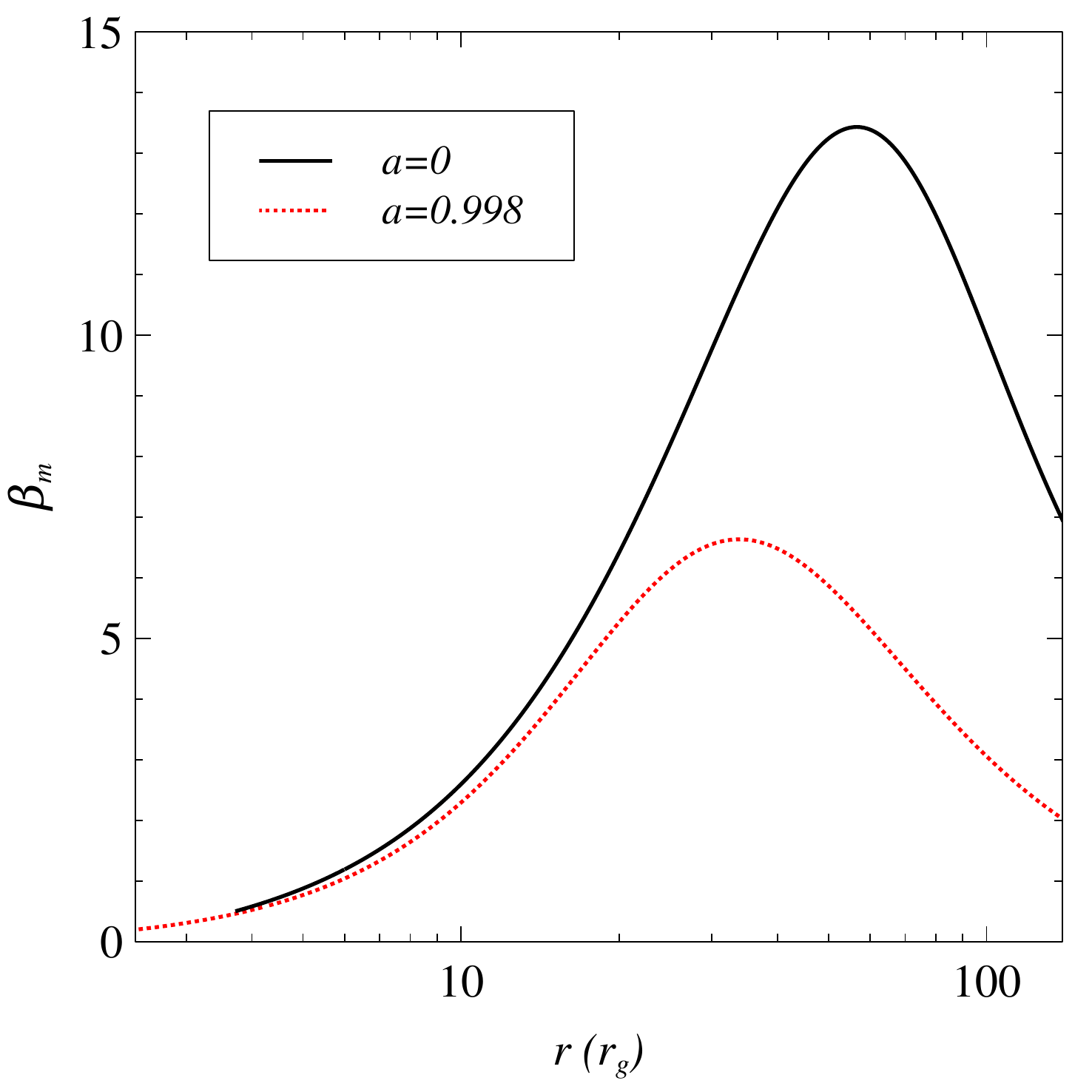}\caption{Variation of plasma-$\beta$ parameter as a function of $r$. The model parameters are same as in Fig~\ref{fig:dynamics}. 
	} 
	\label{fig:plasma_beta}
\end{figure}
\begin{figure}
	%\center
	\includegraphics[width=\columnwidth]{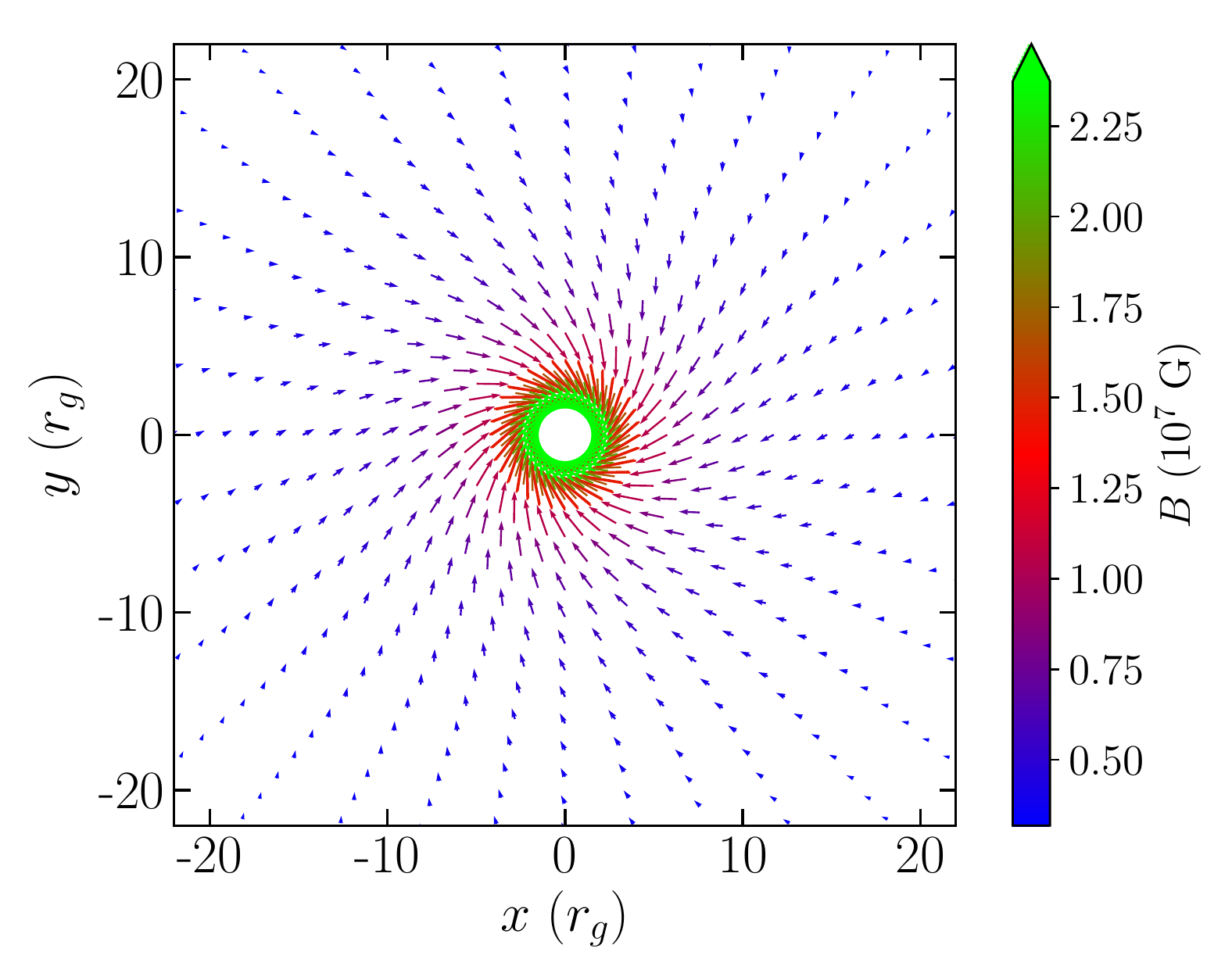}\caption{The nature of magnetic field vectors in the $x-y$ plane of the accretion flow around a rapidly rotating stellar-mass BH. The model parameters are same as in Fig~\ref{fig:dynamics}. The arrow size is normalized with a factor $10^{7}$ here to scale the field value with Cartesian-coordinate value, and the color bar indicates the actual magnetic field strength.
	} 
	\label{fig:field_line}
\end{figure}

The details of the magnetic field geometry and different stress components are shown in Fig.~\ref{fig:field}. Figs.~\ref{fig:field}$(a)$ and $(b)$ indicate the variations of different field components for non-rotating and rapidly rotating BHs respectively. Spinning BHs can sustain more magnetic fields in comparison with non-rotating BHs. The magnetic field strength near the event horizon is of the order of $|B|\sim 8.74\times 10^{6}$ G when $a=0$, and $|B|\sim 2.67\times 10^{7}$ G when $a=0.998$ for a $20 \ M_{\odot}$ BH. Figs.~\ref{fig:field}$(c)$ and $(d)$ show the relative magnitude of different magnetic field components with respect to the total field for $a=0$ and $a=0.998$ respectively. Initially, very far away from the BH, the disc is mainly vertical poloidal field dominated. As matter drags inward, the advection of both poloidal and toroidal fields happens. 
The profiles for the radial poloidal magnetic field and the toroidal magnetic field differ significantly for these two different spin configurations $a=0$ and $a=0.998$ of the BH. The azimuthal component of the magnetic field is strongly linked with the angular momentum profile of the matter on the disc. As mentioned in Fig. \ref{fig:dynamics}$(b)$, the disc angular momentum for a non-spinning BH is more than that for a fast-spinning BH almost throughout the disc. As a result, the nature of the azimuthal and radial components of the magnetic field behaves accordingly for these two spin configurations of the BH, and, hence, angular momentum profiles of the disc.
Figs.~\ref{fig:field}$(e)$ and $(f)$ show the ratio of the different components of magnetic to viscous stresses for $a=0$ and $a=0.998$ respectively. Two main components, $r\phi$, and $\phi z$ reveal the radial and vertical transports, respectively. These profiles signify the dominant nature of the magnetic stress over viscous one, and this should be the case in any magnetically dominated accretion flows. The plasma-$\beta$ $(\beta_{m})$ parameter for both non-spinning and rapidly spinning BHs is shown in Fig.~\ref{fig:plasma_beta}. The rapidly spinning BHs can sustain more magnetic fields compared to the non-spinning BHs. It makes the value of the plasma-$\beta$ parameter smaller for spinning BHs. Also, such low $\beta_{m}$ value (near or below $10$) infers strongly magnetized flows for BH accretion. The nature of the magnetic field vectors is visualized in 2D-plane $(x-y)$ for a rapidly-spinning BH, as shown in Fig.~\ref{fig:field_line}. The arrow size is normalized with a factor $10^{7}$ here to scale the field value with Cartesian-coordinate value, and the color bar indicates the actual magnetic field strength near such rotating stellar-mass BHs.
\vspace{1cm}
\subsection{Disc dynamics: supermassive black holes}

\begin{figure*}
	\center
	\includegraphics[width=16 cm]{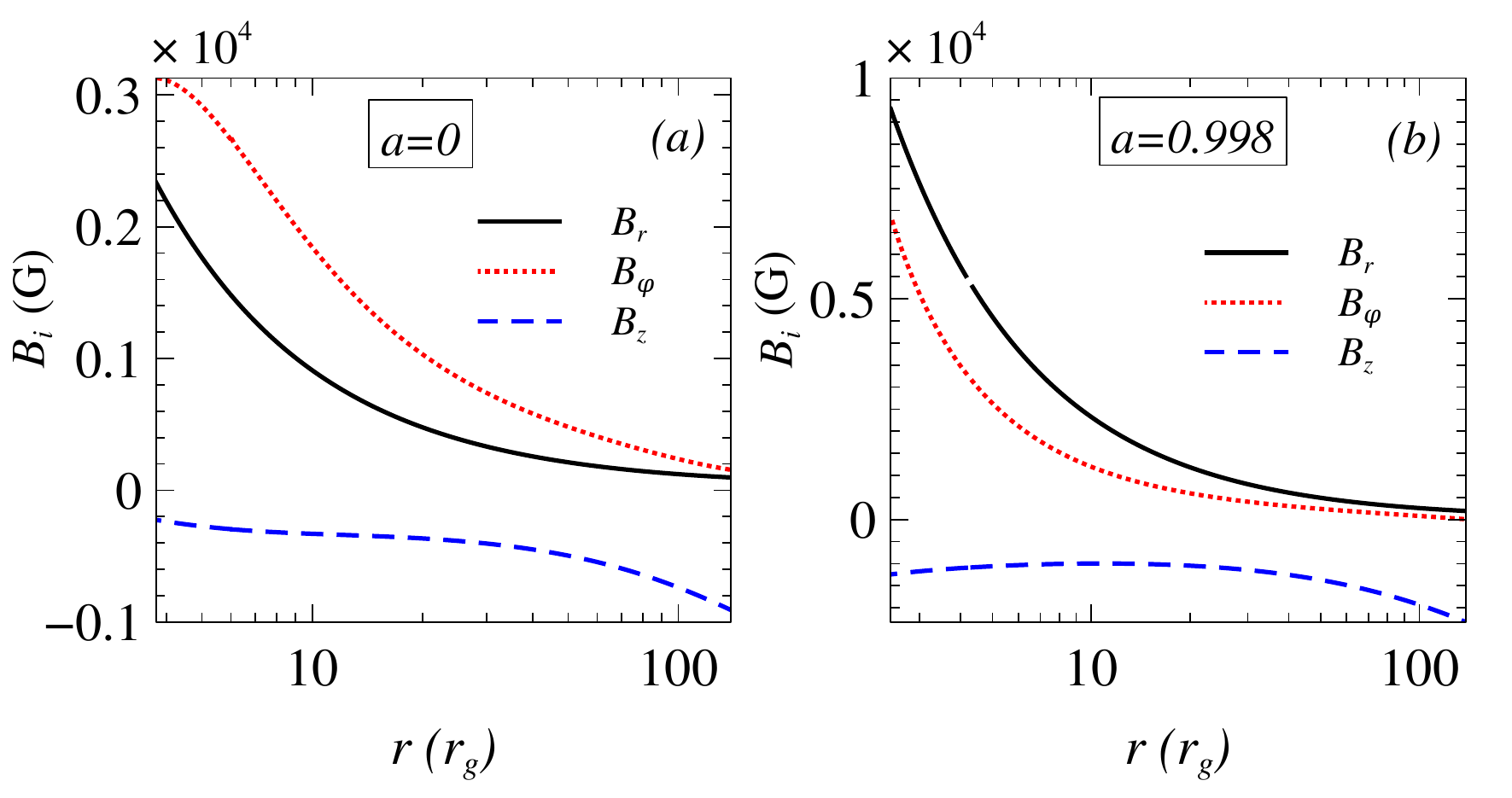}\caption{Variations of magnetic field components for $(a)$ $a=0$, and $(b)$ $a=0.998$, as functions of radial coordinate. The model parameters are $M_{\text{BH}}=10^{8} M_{\odot}$, $\dot{m}=0.05$, and, $\alpha = 0.01$. 
	} 
	\label{fig:field_supermassive}
\end{figure*}

The solutions for magnetized disc-outflow symbiosis around supermassive BHs with mass $10^{8} M_{\odot}$ keeping other model parameters exactly the same as in the case of stellar-mass BHs are shown in Fig.~\ref{fig:field_supermassive}. The flow parameters, like, Mach number, specific angular momentum, outflow velocity, and medium sound speed, reveal the same behaviour to that of stellar-mass BHs, as shown in Fig.~\ref{fig:dynamics}, hence are not depicted again. This is because the magnetized, advective, disc-outflow symbiotic model is effectively scale-free when the physical variables are written in terms of relevant fiducial parameters (Schwarzschild radius, Eddington accretion rate, light velocity, for examples). It makes some quantities, like, radial and vertical velocities, specific angular momentum, sound speed are essentially independent of the BH mass for such advective flows. However, not all the features are similar. Some physical quantities, like, density, magnetic field strength, pressure, differ quite largely with the mass of the BH, though their profiles appear similar. The variations of different magnetic field components for non-rotating, as well as, fast-rotating supermassive BHs are shown in Fig.~\ref{fig:field_supermassive}$(a)$ and $(b)$ respectively. Since the magnetic field strength $(B)$ varies as $(M_{\text{BH}}/M_{\odot})^{-1/2}$, the $B$ value drops here in comparison with stellar mass BHs, as shown in Fig.~\ref{fig:field}. Near the event horizon of the BH, the magnetic field strength is of the order of $|B|\sim 3.91\times 10^{3}$ G when $a=0$, and $|B|\sim 1.16\times 10^{4}$ G when $a=0.998$, for a $10^{8}  M_{\odot}$ BH. The relative magnitude of different magnetic field components with respect to the total field for $a=0$ and $a=0.998$ are the same in comparison with the respective stellar-mass BHs. This is because of the scale-free nature with respect to the mass of the BH. The stress ratios of different components of magnetic to viscous ones also show similar behaviour as in the case of stellar-mass BHs. This is expected because density and pressure vary as $(M_{\text{BH}}/M_{\odot})^{-1}$, and hence stress ratios remain independent of the BH mass.

\subsection{Energetics of the magnetized accretion process}

We compute the energetics of the magnetized accretion induced outflow for non-rotating, as well as, rapidly rotating BHs spanning from stellar-mass to supermassive scales. The disc luminosity can be computed from the cooling mechanisms and can be defined as
\begin{equation}
L=\int \left(\int_{0}^{h} Q^{-}4\pi r \ dz \right)\ dr. \label{eq:luminosity}
\end{equation}
\begin{figure*}
	\center
	\includegraphics[width=16 cm]{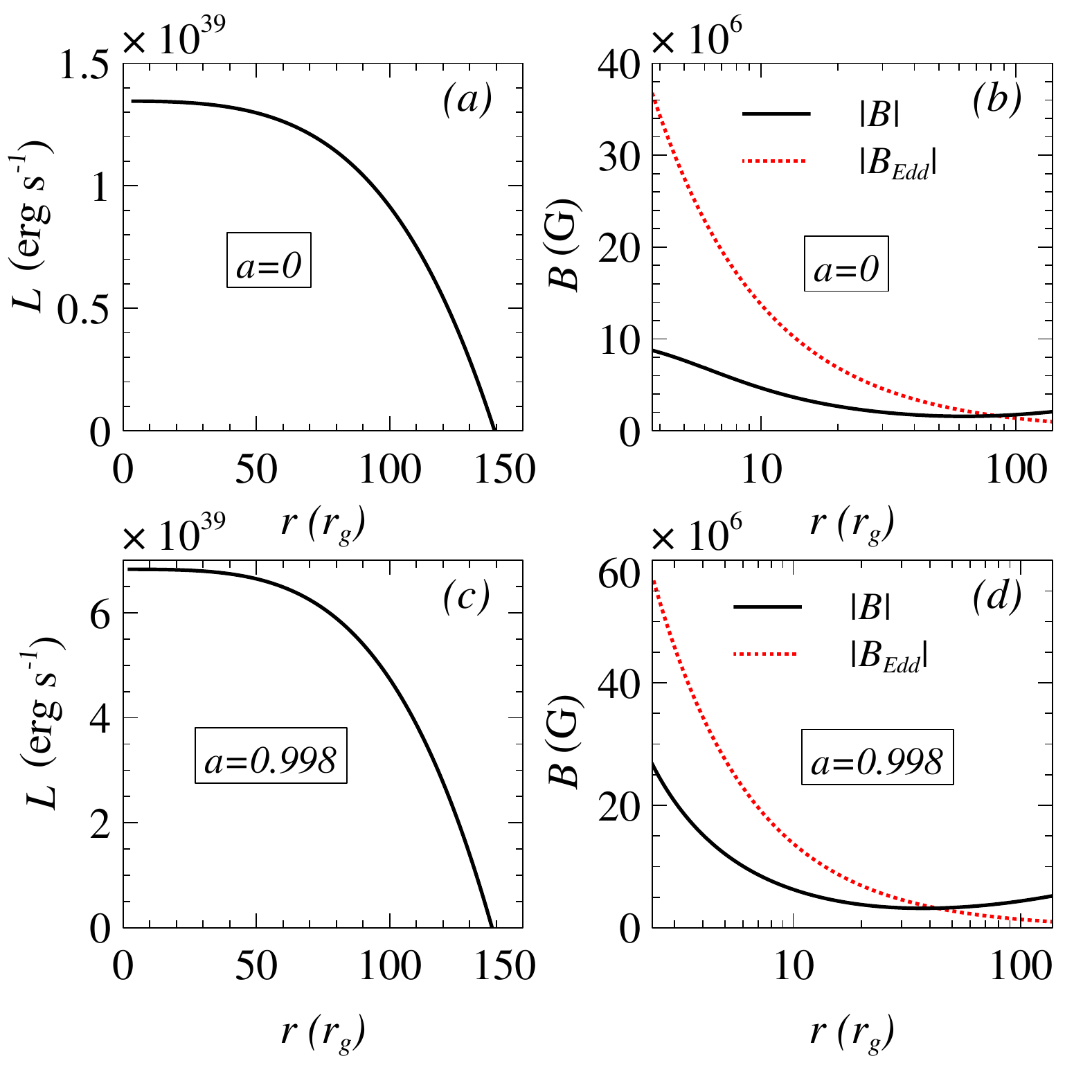}\caption{Variations of $(a)$ disc luminosity, and $(b)$ magnetic field, for a non-rotating stellar mass BH. The black solid line represents the magnetic field profile of the disc, whereas red dotted line represents the corresponding Eddington magnetic field limit. $(c)$, and $(d)$ depict the same as $(a)$, and $(b)$ respectively, except for $a=0.998$. The model parameters are same as in Fig~\ref{fig:dynamics}. 
	} 
	\label{fig:stellar_luminosity}
\end{figure*}
The variation of disc luminosity for a stellar mass BH is shown in Fig.~\ref{fig:stellar_luminosity}. The luminosity at any arbitrary $r$ is computed by integrating (as given in equation~\ref{eq:luminosity}) from outer disc radius to that corresponding $r$. The total luminosity is basically integration over whole disc from outer radius to inner one and, hence, the total luminosity is the value corresponding to that at the inner radius. Figs.~\ref{fig:stellar_luminosity}$(a)$ and $(c)$ show the variations of luminosity for $a=0$ and $a=0.998$ respectively. For stellar-mass BH of mass $M_{BH}=20 M_{\odot}$ with total mass accretion rate $\dot{m}=0.05$, the maximum attainable disc luminosity integrated over whole disc is $L\sim 1.34\times 10^{39}$ erg s$^{-1}$ when $a=0$ and $L\sim 6.83\times 10^{39}$ erg s$^{-1}$ when $a=0.998$. This is because of the fact that the rapidly spinning BHs can sustain more magnetic field compared to non-rotating BHs as given in Figs.~\ref{fig:stellar_luminosity}$(b)$ and $(d)$.
One question may automatically arise: how can such large luminosity possible? The answer is the presence of externally generated large scale strong magnetic field and its configurations.
The red dotted lines in Figs.~\ref{fig:stellar_luminosity}$(b)$ and $(d)$ show the variation of the Eddington magnetic field limit $(B_{\text{Edd}})$ for $20 M_{\odot}$ BH. Following \cite{2010PhyU...53.1199B}, the estimate of such $B_{\text{Edd}}$ is based on the simple assumption that the luminosity associated with the magnetic energy density is comparable to the Eddington luminosity, given by equation (\ref{eq:L_Edd}). This can be expressed as
\begin{equation*}
	c \frac{B_{\text{Edd}}^{2}}{8\pi}4\pi r^{2}=L_{\text{Edd}},
\end{equation*}
which further can be simplified as
\begin{equation}
B_{\text{Edd}} \approx \frac{6.15\times 10^{8}}{r} \left(\frac{M_{\text{BH}}}{M_{\odot}} \right)^{-1/2} \ \text{G}. \label{eq:B_Edd}
\end{equation}
The magnetic field near the outer disc region is large enough and even more than the corresponding Eddington limit $B_{\text{Edd}}$ at that zone. Near the BH, the magnetic field maintains its Eddington limit. Such field configurations enhance the synchrotron and SSC processes and, hence, help to achieve very large luminosity.
\begin{figure*}
	\center
	\includegraphics[width=16 cm]{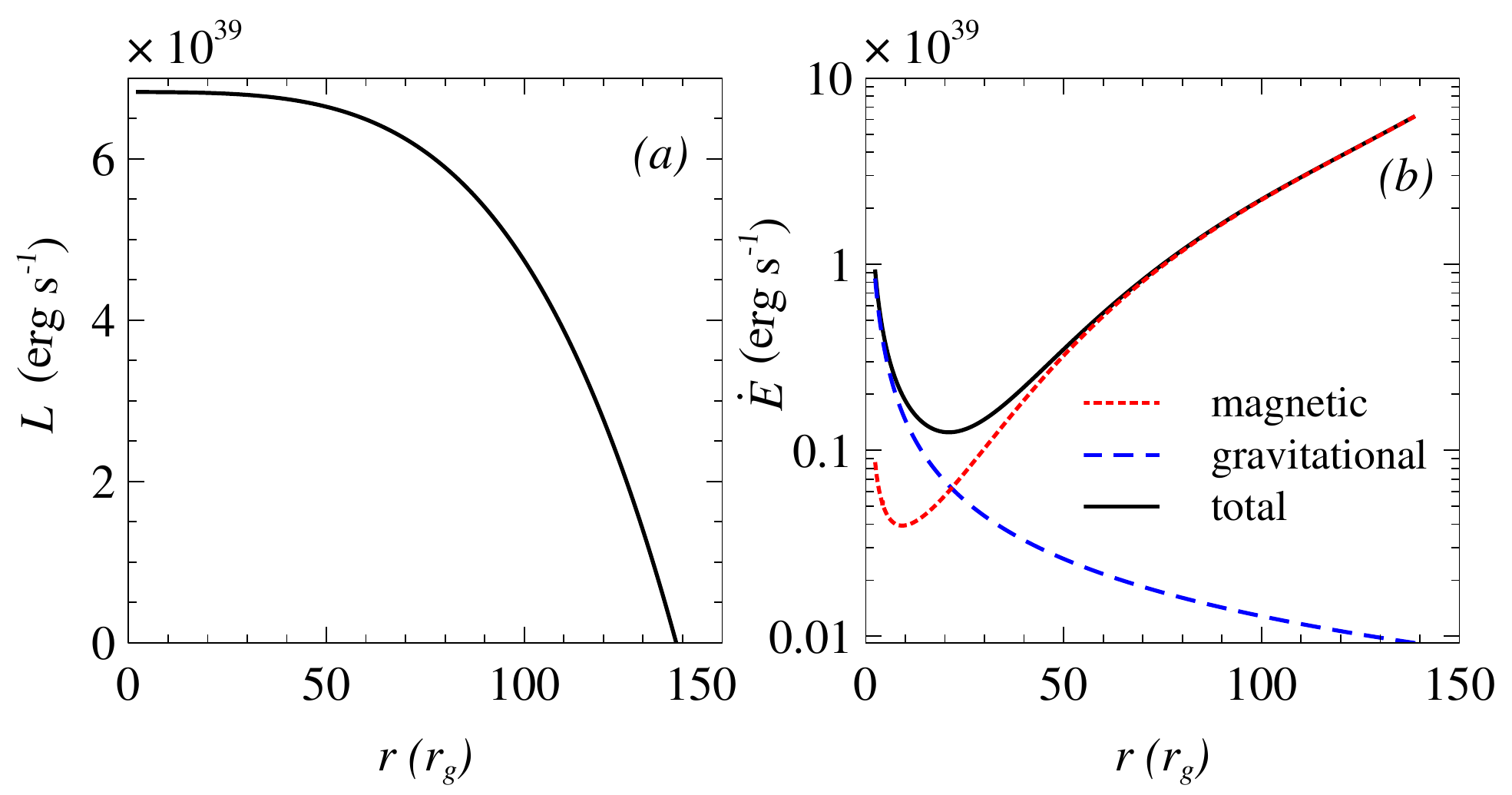}\caption{Variations of $(a)$ disc luminosity, and $(b)$ rate of change of energy from gravitational potential (magnitude here), magnetic counterpart, and total, as functions of radial coordinate. This is basically to address the energy conservation principle for the case as shown in Figs.~\ref{fig:stellar_luminosity}$(c)$ and $(d)$. The model parameters are $M_{\text{BH}}=20 M_{\odot}$, $\dot{m}=0.05$, and, $\alpha = 0.01$. 
	} 
	\label{fig:energy}
\end{figure*}
One of the most important points is the energy conservation principle, which is explained here based on Fig.~\ref{fig:energy}. The maximum gravitational potential energy per unit time is $|\dot{E}_{G}| \sim 8.3\times 10^{38}$ erg s$^{-1}$, based on a $20M_{\odot}$ BH with mass accretion rate $\dot{m}=0.05$. However the maximum energy associated with the magnetic field per unit time $(\dot{E}_{B}=v \frac{B^{2}}{8\pi}4\pi rh$, where $h$ is  the  scale  height  of the flow) for this configuration is $\dot{E}_{B} \sim 6.2\times 10^{39}$ erg s$^{-1}$ as shown in Fig.~\ref{fig:energy}$(b)$. Therefore $L<|\dot{E}_{G}|+\dot{E}_{B}$ as shown in Fig.~\ref{fig:energy}$(a)$. This magnetic energy decreases through synchrotron and SSC cooling processes as matter drags towards the central BH with increasing gravitational potential. Near the plunging region, most of the magnetic energy is used off, and everything is controlled by the gravitational potential of the BH. Therefore, far away, it is the magnetic energy, and close to the BH, the gravitational potential energy controlling the energy budget.
Note that the advections, radial and vertical, play a very important role in maintaining the equilibrium solutions in the presence of such large-scale strong magnetic fields. Unlike the Keplerian disc, matter in the advective regime does not sustain at a radius, hence the disc does not disappear by the buoyancy effect or otherwise. Note further that unlike most of the existing approaches, here we consider non-zero vertical velocity $v_{z}$ also. Hence, the conventional vertical static equilibrium is not coming in this picture. 
\begin{figure*}
	\center
	\includegraphics[width=16 cm]{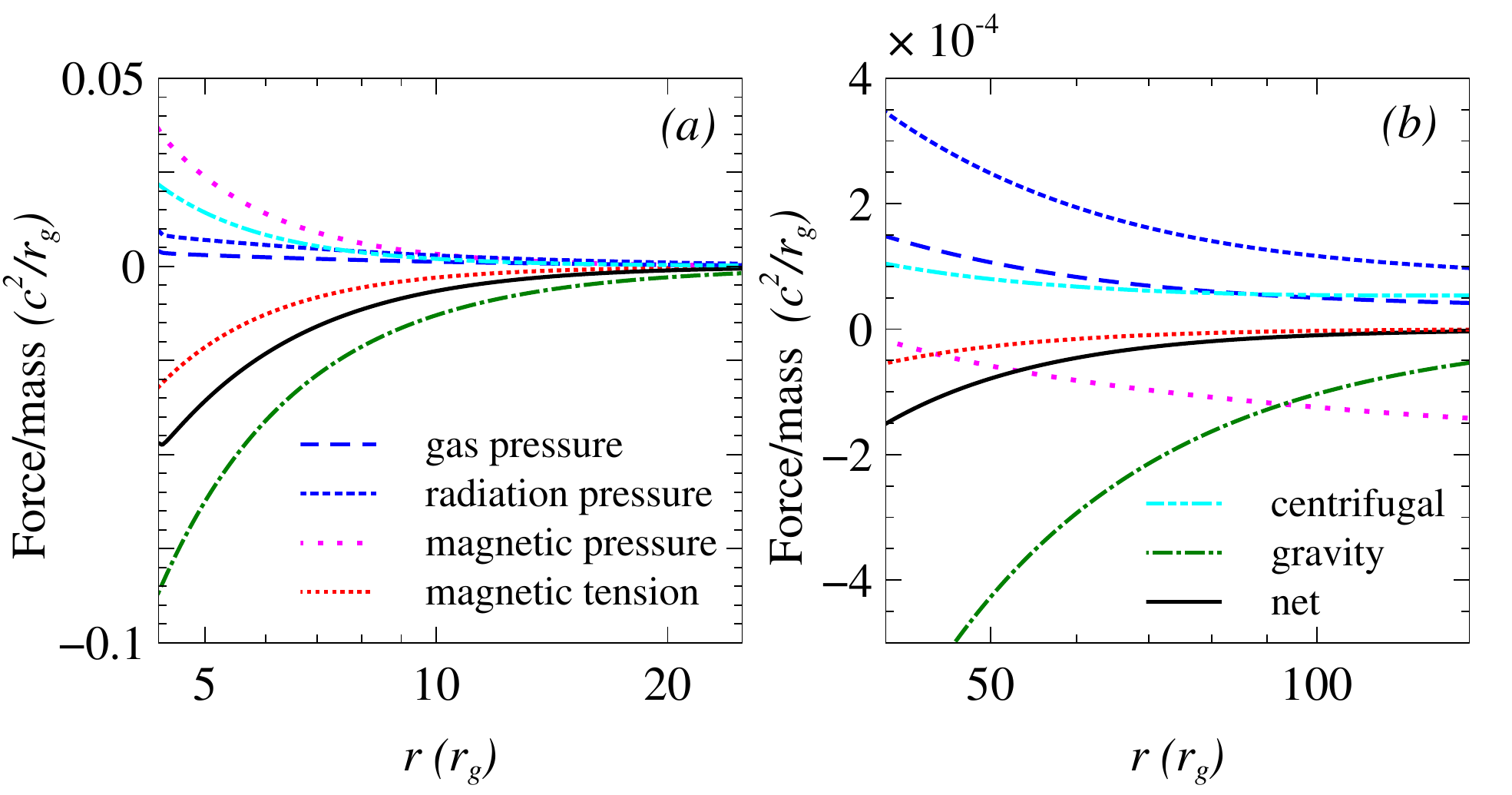}\caption{Variations of different force terms per unit mass as functions of radial coordinate for two different disc locations: $(a)$ near the BH, and $(b)$ outer disc region. This is basically to address the different acceleration terms for the case shown in Figs.~\ref{fig:stellar_luminosity}$(c)$ and $(d)$.  
	} 
	\label{fig:force}
\end{figure*}
The other important point is that the Lorentz force introduces a gradient of pressure $B^{2}/8\pi$ and a tension term $(\mathbf{B}.\mathbf{\nabla})\mathbf{B}$ along magnetic field lines. The magnetic pressure term generally acts against gravitational force, behaves as a normal fluid pressure. However, the magnetic tension term mostly behaves oppositely, effectively like a negative pressure supporting gravity. Since the luminosity in our system is more than $L_{\text{Edd}}$, the gradient of radiation pressure is also significant to hold in the system. To capture the effect of forces associated with both the strong magnetic fields and the radiation pressure, we need to look at the radial momentum balance equation carefully. Neglecting the term associated with $W_{rz}$, we can rewrite equation (\ref{rad_momentum}) as
	\begin{multline}
	v_{r}\frac{\partial v_{r}}{\partial r}+v_{z}\frac{\partial v_{r}}{\partial z}= \\
	\frac{\lambda ^{2}}{r^{3}}-\frac{1}{\rho}\frac{\partial p_{g}}{\partial r}-\frac{1}{\rho}\frac{\partial p_{r}}{\partial r}-F-\frac{1}{\rho}\frac{\partial}{\partial r}\left(\frac{B^{2}}{8\pi}\right)+\frac{\left[(\mathbf{B}.\nabla) \mathbf{B}\right]_{r}}{4\pi \rho}. \label{eq:force}
	\end{multline} 
The advection terms are in the L.H.S. of the equation (\ref{eq:force}), whereas R.H.S. of this equation involves all the acceleration terms: the centrifugal, the gradient of gas pressure, the gradient of radiation pressure, gravitational acceleration, the gradient of magnetic pressure, and the effect due to magnetic tension respectively. The evolution of individual force/acceleration terms are shown in Figure~\ref{fig:force}. Here, the net force indicates a sum over all the forces. The second inward force, in addition to gravity, is associated with the magnetic fields. The magnetic tension always supports the gravity. In the outer disc region, however, the gradient of magnetic pressure also operates inward direction for the given geometry. This is because, once the magnetic field is supplied therein, the synchrotron cooling starts operating. Hence, the magnetic field strength continuously drops, resulting in matter drags inward in this particular zone. Most of the magnetic energy is used off, and also the integrated disc-luminosity is enhanced and getting saturated in this region, as given by Figure \ref{fig:energy}. The overall net force is negative throughout the disc. The contributions from magnetic pressure and tension depend on the field geometry. Hence the necessary condition to maintain equilibrium solution is the anisotropic nature of the magnetic field through the magnetic tension and the non-negligible advections, as given by equations~(\ref{rad_momentum}) and (\ref{vertical_momentum}). It is the field geometry, in addition to the field strength, which is exploited in order to achieve our result, as similarly was done earlier by \cite{1976MNRAS.175..395B} in the context of channelled accretion flows around neutron stars.

\begin{figure*}
	\center
	\includegraphics[width=16 cm]{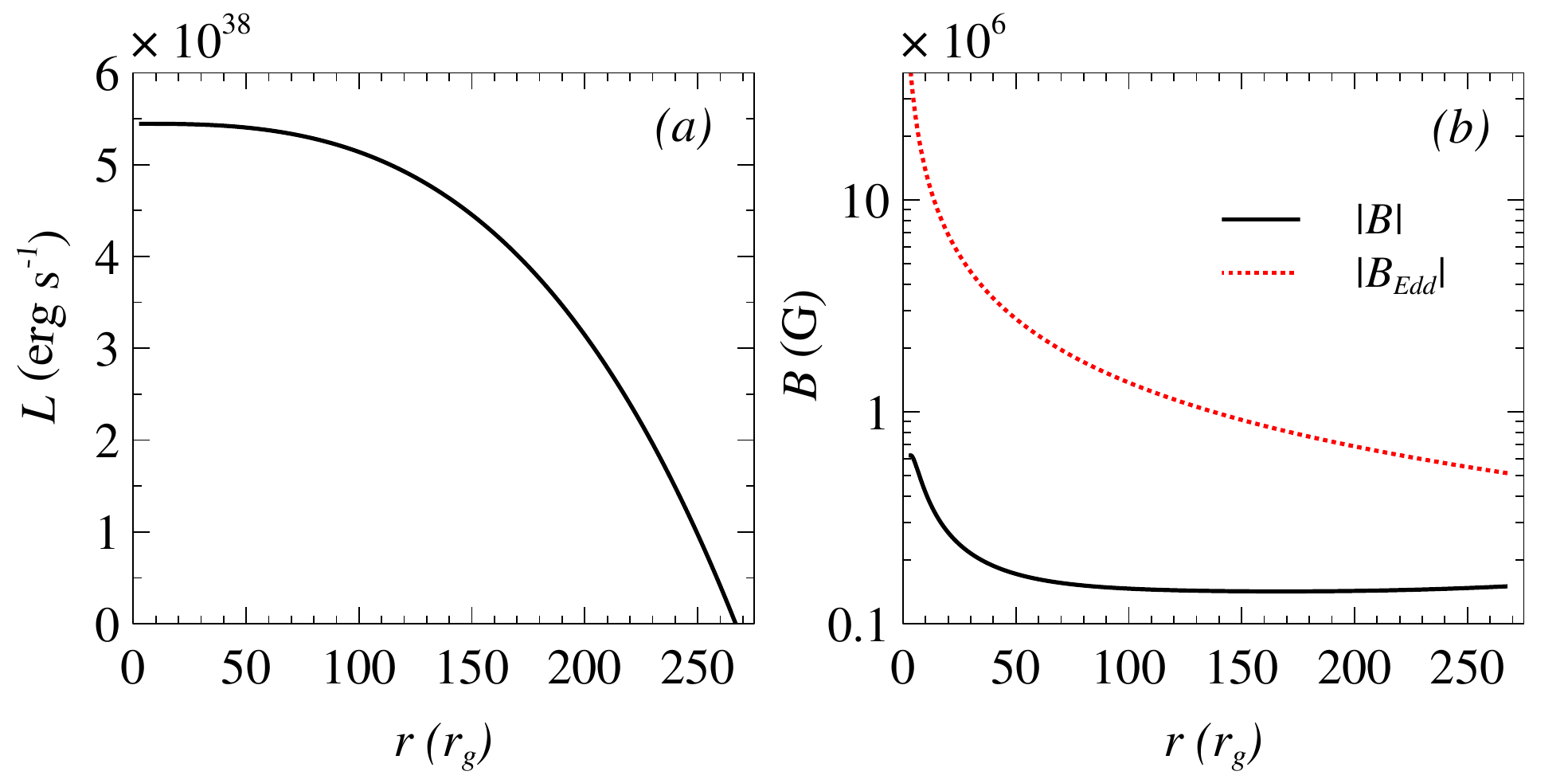}\caption{Variations of $(a)$ disc luminosity, and $(b)$ magnetic field, for a non-rotating stellar mass BH for the case of weak fields than that considered in Fig.~\ref{fig:stellar_luminosity}. The black solid line represents the magnetic field profile of the disc, whereas red dotted line represents the corresponding Eddington magnetic field limit. The model parameters are same as in Fig~\ref{fig:stellar_luminosity}, except for weaker magnetic field.
	} 
	\label{fig:stellar_luminosity2}
\end{figure*}
If magnetic field strength drops due to less supply of magnetic field and goes down below the corresponding $B_{\text{Edd}}$ limit throughout the flow, then synchrotron, as well as, SSC cooling also reduce accordingly. This accretion environment certainly can not achieve very large luminosity. This is shown in Fig.~\ref{fig:stellar_luminosity2} for the case of non-rotating stellar-mass BHs with mass $20 M_{\odot}$, keeping other parameters same as shown in Fig.~\ref{fig:stellar_luminosity}. The maximum attainable disc luminosity integrated over the whole disc here is $L\sim 5.44\times 10^{38}$ erg s$^{-1}$. Fig.~\ref{fig:stellar_luminosity2}$(b)$ shows the variation of the magnetic field strength along with its corresponding Eddington limit. Near the event horizon of the BH, the magnetic field strength is of the order of $|B|\sim 6.2288\times 10^{5}$ G, and it is one order less than the case as shown in Fig.~\ref{fig:stellar_luminosity}. The field strength here is quite below the corresponding $B_{\text{Edd}}$ limit throughout the flow. Hence the magnetic field strength is playing a significant role in achieving enormous luminosity.

\begin{figure*}
	\center
	\includegraphics[width=16 cm]{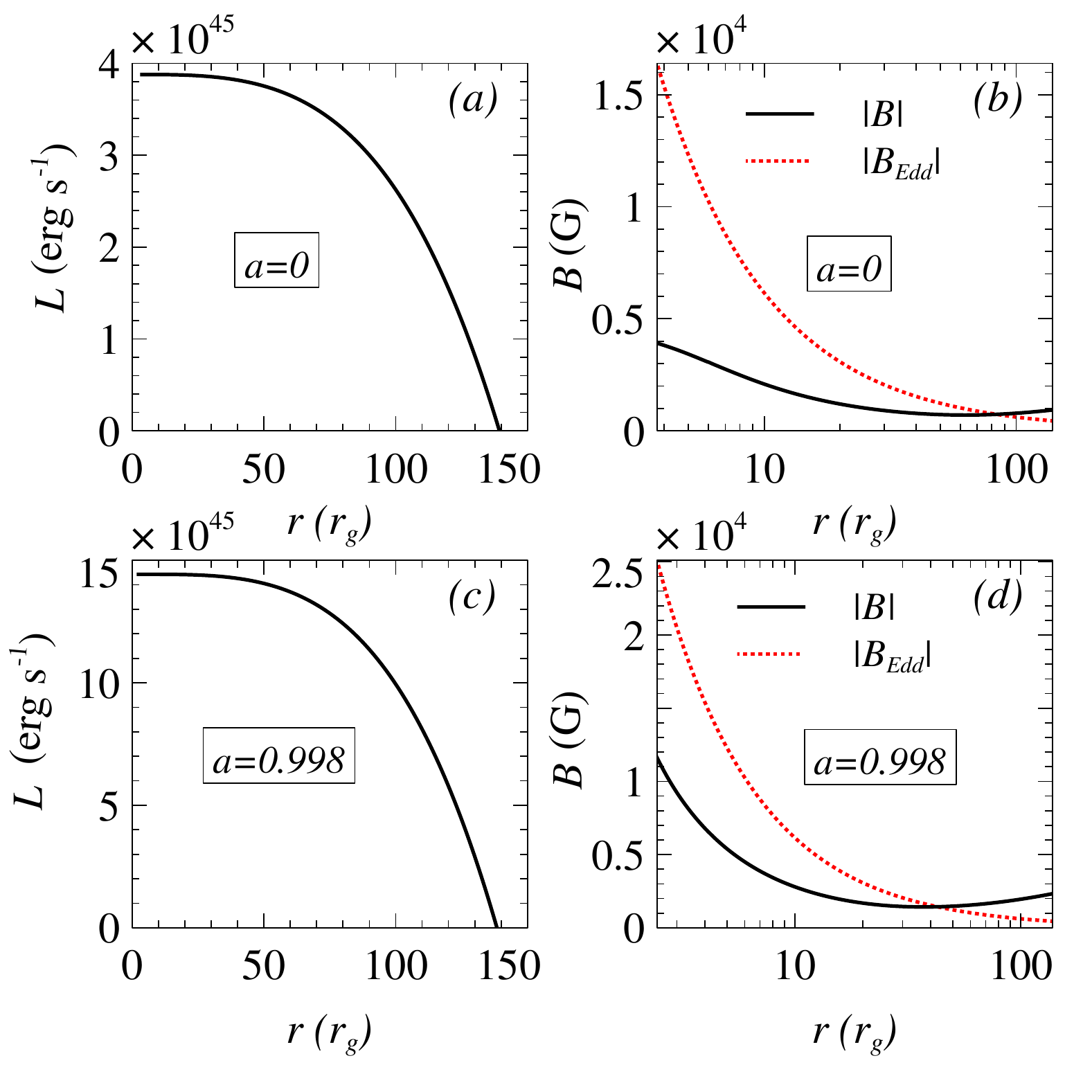}\caption{Variations of $(a)$ disc luminosity, and $(b)$ magnetic field profile for a non-rotating supermassive BH. The black solid line represents the magnetic field profile of the disc, whereas red dotted line represents the corresponding Eddington magnetic field limit. $(c)$, and $(d)$ depict the same as $(a)$, and $(b)$ respectively, except for $a=0.998$. The model parameters are same as in Fig~\ref{fig:dynamics}, except $M_{\text{BH}}=10^{8} M_{\odot}$.
	} 
	\label{fig:supermassive_luminosity}
\end{figure*}
The accretion disc luminosity and the corresponding magnetic field strength for a supermassive BH of mass $M_{BH}=10^{8}  M_{\odot}$ with $\dot{m}=0.05$ are shown in Fig.~\ref{fig:supermassive_luminosity}. The luminosity is strongly BH mass dependent. For instance, the synchrotron emission is self-absorbed below the critical frequency $\nu_{c}$ and it is radiated at $\nu_{c}$ which is proportional to the magnetic field strength $B$ (equation \ref{eq:cut_off_frequency}) and hence this critical frequency itself varies as $(M_{\text{BH}}/M_{\odot})^{-1/2}$. The maximum attainable disc luminosity integrated over whole disc is $L\sim 3.88\times 10^{45}$ erg s$^{-1}$ when $a=0$ and $L\sim 1.44\times 10^{46}$ erg s$^{-1}$ when $a=0.998$. As described for the case of stellar-mass BH, this is because of the fact that the rapidly spinning BHs can sustain more magnetic field compared to non-rotating BHs and therefore enhance cooling process through synchrotron radiation and SSC process.

\section{OBSERVATIONAL IMPLEMENTATIONS} \label{sec:observations}

Galactic BH X-ray binaries show large varieties in their X-ray spectral states, namely, quiescent, low/hard, intermediate, high/soft, and very high \citep{2006ARA&A..44...49R, 2004MNRAS.355.1105F}. It is believed that these different spectral states correspond to different accretion geometries, as mentioned before. In this paper, we focus on the hard-state BH sources with BH mass ranging from stellar mass to supermassive scales. The so-called hard states are generally observed at low X-ray luminosities $(L<0.01L_{\text{Edd}})$, and their X-ray spectra are well explained by a power-law component with photon index $\Gamma \sim 1.4$ to 1.8. However, the observations of hard state BH sources often reach higher luminosities. For example, the BHC GX 339-4 achieves luminosity up to $0.3 L_{\text{Edd}}$ in its hard state \citep{2004MNRAS.351..791Z}. 
%The RXTE observations for BH binary XTE J1550-564 reaches luminosity $\sim 20\% L_{\text{Edd}}$ during its 1998-1999 outburst \citep{2000ApJ...544..993S}. 
Apart from such BH binaries, a large fraction of ULXs are even more luminous in their hard power-law dominated states. Some of these ULXs are listed in Table \ref{tab:ulx}. The true nature of such observations remains mysterious over the decades. Here we address that our highly magnetized, advective, optically thin disc-outflow symbiotic model can achieve these large luminosities. Fig. \ref{fig:stellar_luminosity} indicates that maximum attainable luminosity integrated over the disc is $L\sim 6.83\times 10^{39}$ erg s$^{-1}$ for a rapidly spinning stellar-mass BH of mass $M_{BH}=20 M_{\odot}$ with mass accretion rate $\dot{m}=0.05$. Hence such a magnetically dominated advective accretion process can easily explain these long-standing issues. 

On the other hand, the observations of supermassive BHs also show unusual large luminosities in their hard spectral states. For example, some high synchrotron peak (HSP) BL Lac objects appear very luminous in their hard power-law dominated states. Some of these sources are listed in Table \ref{tab:BL_Lac} based on the observations of $Fermi$ second catalog of active galactic nuclei \citep{2011ApJ...743..171A}. As shown in Fig. \ref{fig:supermassive_luminosity} that our magnetically dominated disc-outflow symbiotic model can reach luminosity $L\sim 1.44\times 10^{46}$ erg s$^{-1}$ for a rapidly spinning supermassive BH of mass $M_{BH}=10^{8} M_{\odot}$ with mass accretion rate $\dot{m}=0.05$. 
 
\begin{table}
	\centering
	\caption{Some hard-state ULX sources.}
	\label{tab:ulx}
	%\begin{threeparttable}
		\begin{tabular}{lccr} % four columns, alignment for each
			\hline
			Source & $\Gamma$ & $L_{0.3-10 \ keV}$  & Ref.\\
			&          & $(10^{40} \ erg \ s^{-1})$ & \\
			\hline       
			NGC 1365 X1 & $1.74^{+0.12}_{-0.11}$ & 2.8 & 1\\
			& $1.80^{+0.04}_{-0.05}$ & 0.53 &  \\\\
			
			NGC 1365 X2 & $1.23^{+0.25}_{-0.19}$ & 3.7 &  \\
			& $1.13^{+0.09}_{-0.10}$ & 0.15 &  \\\\
			
			Holmberg IX X-1 & $1.9^{+0.1}_{-0.02}$ & 1.0 & 2\\\\
			
			NGC 5775 X1 & $1.8^{+0.3}_{-0.2}$ & 7.5 & 3\\\\
			
			NGC 3628 X1 & $1.8^{+0.2}_{-0.2}$ & 1.1 & 4\\\\       
			
			M99 X1 & $1.7^{+0.1}_{-0.1}$ & 1.9 & 5\\\\
			
			M82 X42.3+59 & $1.44^{+0.09}_{-0.09}$ & 1.13 & 6\\
			& $1.33^{+0.13}_{-0.13}$ & 1.51 &  \\\\
			
			Antennae X-11 & $1.76^{+0.05}_{-0.05}$ & 2.11 & 7\\
			& $1.68^{+0.06}_{-0.06}$ & 1.38 &  \\\\
			
			Antennae X-16 & $1.35^{+0.03}_{-0.04}$ & 1.82 &  \\
			& $1.2^{+0.14}_{-0.10}$ & 0.90 &  \\\\
			
			Antennae X-42 & $1.73^{+0.10}_{-0.11}$ & 0.96 &  \\
			& $1.66^{+0.05}_{-0.06}$ & 1.00 &  \\\\
			
			Antennae X-44 & $1.74^{+0.04}_{-0.04}$ & 1.28 &  \\
			& $1.63^{+0.09}_{-0.09}$ & 1.48 &  \\
			
			%\hline           
			\hline
		\end{tabular}
		\begin{tablenotes}
			\item References: $(1)$ \cite{2009ApJ...695.1614S}; $(2)$ \cite{2009ApJ...702.1679K}; $(3)$ \cite{2008MNRAS.390...59L}; $(4)$ \cite{2001ApJ...560..707S}; $(5)$ \cite{2006MNRAS.372.1531S}; $(6)$ \cite{2010ApJ...710L.137F}; $(7)$ \cite{2009ApJ...696.1712F}.
		\end{tablenotes}
	%\end{threeparttable}
\end{table}

\begin{table}
	\centering
	\caption{BL Lac objects in a hard power-law dominated state based on the observations of $Fermi$ second catalog of active galactic nuclei \citep{2011ApJ...743..171A}.}
	\label{tab:BL_Lac}
	%\begin{threeparttable}
	\begin{tabular}{lccr} % four columns, alignment for each
		\hline
		Source & $\Gamma$ & $L_{1-100 \ \text{GeV}}$  \\
		&          & $(10^{46} \ \text{erg \ $s^{-1}$})$ \\
		\hline       
		KUV 00311-1938 & 1.758 & 4.44 \\       
		
		PKS 0301-243 & 1.938 & 1.10 \\
		
		PKS 0447-439 & 1.855 & 1.21 \\
		
		1ES 0502+675 & 1.489 & 1.83 \\
		
		1H 1013+498 & 1.723 & 1.05 \\

		B3 1307+433 & 1.839 & 3.16 \\
		
		1H 1515+660 & 1.665 & 1.49 \\
            
		\hline
	\end{tabular}
	%\begin{tablenotes}
	%	\item References: $(1)$ \cite{2001ApJ...560..707S}; $(2)$ \cite{2006MNRAS.372.1531S}; $(3)$ \cite{2009ApJ...696.1712F}; $(4)$ \cite{2009ApJ...702.1679K}; $(5)$ \cite{2009ApJ...695.1614S}; $(6)$ \cite{2010ApJ...710L.137F};
	%\end{tablenotes}
	%\end{threeparttable}
\end{table}

\section{DISCUSSION} \label{sec:discussion}

In this $2.5-$dimensional magnetized, viscous, advective disc-outflow symbiotic model, we address the role of large-scale strong magnetic fields in the formation of strong outflows and the enhancement of synchrotron and SSC cooling. Below we discuss some important aspects of this framework.

First, what could be the possible origin of large-scale strong magnetic fields in an accretion disc? The accretion disc problems have been studied widely in the presence of small-scale fields, as well as large-scale fields. The small-scale field may be generated locally. Some seed magnetic fields can generate from zero initial field conditions via the Biermann Battery mechanism \citep{biermann1950biermann}. To operate this mechanism, a non-parallel gradients of density and temperature profiles are required, which is very common in the accretion environment. Also, the MHD dynamo process may generate a small-scale magnetic field locally \citep{1995ApJ...446..741B}. However, the origin of large-scale magnetic fields is not clear yet. One possibility may be the externally generated fields. The interstellar medium or the companion star may supply such fields, which further push towards the central BH by the continuous accretion pressure and may enhance due to flux freezing \citep{1974Ap&SS..28...45B}.  

Second, is there any upper bound to the amount of magnetic flux to thread the disc and BH? Any magnetized accretion flow causes a certain amount of magnetic flux to thread the disc and BH. Recent observations of radio-loud active galaxies confirm a dynamically dominated magnetic field in the jet launching region based on the correlation of jet magnetic field and accretion disc luminosity \citep{2014Natur.510..126Z}. Also, the unusual large Faraday rotation near the centre of our Galaxy infers a signature of a strong magnetic field near the BH \citep{2013Natur.501..391E}. GRMHD simulations for relativistic jets generally assume highly magnetized plasma at the jet foot-print \citep{2004ApJ...611..977M, 2011MNRAS.418L..79T}. Theoretical models over the decades are trying to correlate the observable quantities with the fundamental properties of the disc and BH. Based on the combined effects of Blandford-Payne \citep{1982MNRAS.199..883B} and Blandford-Znajek \citep{1977MNRAS.179..433B} mechanisms, the jet kinetic power had been computed in terms of the mass and spin of the BH and the magnetic field strength in the vicinity of BH \citep{2010MNRAS.406..975G}. From an estimate of the kinetic power of the relativistic jet \citep{2013MNRAS.431..405R}, the measurement of the synchrotron cooling time \citep{2016A&A...593A..47B} and the observed characteristic frequencies of quasi-periodic oscillations of radiation \citep{2011AstBu..66..320P}, it was suggested that the typical values of the magnetic field near the event horizon is $B \sim 10^{8}$ G for stellar-mass BHs and $B \sim 10^{4}$ G for supermassive BHs. In this context, the Eddington magnetic field limit near the event horizon of a BH, $B_{Edd}=10^{4} \ \text{G} \ \left(\frac{M_{\text{BH}}}{10^{9}M_{\odot}} \right)^{-1/2}$, might be the approximate upper bound to the amount of magnetic field strength what any disc around a BH can sustain \citep{2019MNRAS.482L..24M}, which limit is perfectly viable in our computation. More importantly, this upper limit in our model sets up automatically through the existence of the inner `saddle'-type critical point, as discussed before.

%Third, what are the roles of geometrically thick advective accretion flow here?

Apart from these, our main venture is to explain some unusual observational features based on this magnetized accretion process. One such observation lies 
with ULXs, which are very bright, off-nuclear, X-ray sources with luminosity exceeding the standard Eddington limit for a neutron star (NS) or even a stellar-mass BH. 
There has been a long debate between NS and BH as a central compact object in ULXs. The recent discoveries of coherent pulsations in three ULXs (M82 X-2, \citealt{2014Natur.514..202B}; NGC 7793 P13, \citealt{2016ApJ...831L..14F}; and NGC 5907 ULX-1, \citealt{2017Sci...355..817I}) imply that the compact object in each of these systems is a NS. The accreting magnetized NS can achieve the apparent super-Eddington luminosity either through the column accretion flow \citep{1976MNRAS.175..395B} or through the suppression of the electron scattering cross-section \citep{1971PhRvD...3.2303C}.
The theoretical explanation for large apparent luminosities of BH ULXs is either the existence of the missing class of intermediate-mass black hole  \citep[IMBH;][]{2004ApJ...614L.117M}, or the super-Eddington accretion around a stellar-mass BH \citep{2003ApJ...597..780E}, or beamed emission \citep{2001ApJ...552L.109K}. Indeed, the existence of the missing class of IMBH is still an open question, and also the supporting evidence for the IMBH scenario has been disputed extensively \citep{2006MNRAS.371..673G}. Most arguments support the idea that a significant fraction of ULXs is stellar-mass BH \citep{2002ApJ...568L..97B, 2014Natur.514..198M}. Super-Eddington scenario may explain the ULXs with a steep power law. However, the hidden nature of a large fraction of ULXs with hard-power law state remains mysterious. Exactly at this point, our model comes. The magnetically dominated advective disc-outflow symbiosis around a stellar-mass BH can reach such large luminosity even for sub-Eddington mass accretion rate. The most important criterion is the magnetic field strength, as well as its geometry. Near the transition region from the Keplerian to sub-Keplerian flows, the magnetic field strength is quite larger than the corresponding Eddington limit. This field becomes dynamically dominant near the BH due to continuous advection of the magnetic flux through flux freezing. However, near the BH, the magnetic field well maintains its corresponding Eddington limit. This is a very efficient way to enhance the synchrotron and SSC cooling to reach such a large luminosity. Also, the large toroidal fields exert an immense outward pressure to produce strong outflows in this model. Unlike other magnetically dominated accretion models, here, the advection of both poloidal and toroidal field components takes place.

\section{CONCLUSIONS} \label{sec:conclusion}

We have obtained the semi-analytical solutions for a magnetized, advective, optically thin disc-outflow symbiosis incorporating explicit cooling formalisms. The detailed balance between heating, cooling, and advection is taken care of here. We prescribe the generalized viscous shearing stress in terms of standard $\alpha$-viscosity parameter. However, the components for magnetic shear are at least one order of magnitude larger than that of viscous shear in this formalism. The large scale strong magnetic field removes the angular momentum from the in-falling matter and also helps in the formation of strong outflows. We assume that the energy transfer from ions to electrons occurs through Coulomb coupling, and the radiative cooling processes are bremsstrahlung emission, synchrotron emission, and the corresponding inverse Compton effects. The presence of strong magnetic fields enhances the synchrotron cooling and SSC process via hot electrons. We implement our model to explain the observed unusually large luminosity of BH sources of mass ranging from stellar mass to supermassive scale in their hard power-law dominated states. Particularly we focus on ULXs (listed in Table 1), BH binaries with luminosities up to $\sim 0.1$ to $0.3 L_{\text{Edd}}$ in their hard state (e.g., GX 339-4), ultra-luminous quasars (e.g., PKS 0743-67), and highly luminous BL Lac objects (listed in Table 2). We propose here for the first time in the literature that such bright HSP-BL Lacs with hard spectral signature are magnetically powered sub-Eddington, advective accretors around supermassive BHs. In a similar framework, for the stellar-mass scale, we suggest that ULXs are magnetically dominated, advective, sub-Eddington accretors around rapidly-rotating stellar-mass BHs, neither incorporating the existence of intermediate-mass BHs concept nor with the super-Eddington accretion phenomenon.

Apart from such important aspects of this strongly magnetized advective accretion process, we would like to explore some poorly understood, long-standing questions in this direction in the future. First, throughout our computation, we assume that the energy transfer from ions to electrons is happening due to Coulomb coupling. How do the other non-thermal processes influence the results in such magnetically dominated advective accretion phenomena? Second, in our disc-outflow model, we use pseudo-Newtonian potential, which might not be suitable to extract the rotational energy of a BH. We plan to explore such magnetically dominated advective accretion flows in the framework of GRMHD formalism in the future. Third, how does electron and ion energization occur in such an optically thin, strongly magnetized, advective accretion process? Very recently, \cite{PhysRevLett.122.055101} and \cite{2019JPlPh..85c9003S} initiated to understand the electron and ion heating, coupling, and also the non-thermal particle acceleration mechanism in the case of radiatively inefficient accretion flows around BHs. Fourth, what are the plausible mechanisms to generate large-scale magnetic fields locally in this advective accretion flows, and what is the strength of such generated magnetic fields?

\section*{ACKNOWLEDGEMENTS}

The authors thank Sudeb Ranjan Datta of IISc, Bangalore, for helpful discussions. Further thanks are due to the MNRAS scientific editor handling the manuscript and the anonymous referee for comments and suggestions, which have improved the presentation of the paper. The work was partly supported by the Department of Science and Technology project with Grant No. DSTO/PPH/BMP/1946 (EMR/2017/001226).

%%%%%%%%%%%%%%%%%%%%%%%%%%%%%%%%%%%%%%%%%%%%%%%%%%


%\bibliographystyle{mnras}
%\bibliography{mhd}

%%%%%%%%%%%%%%%%%%%%%%%%%%%%%%%%%%%%%%%%%%%%%%%%%%

%%%%%%%%%%%%%%%%% APPENDICES %%%%%%%%%%%%%%%%%%%%%

%\appendix


%%%%%%%%%%%%%%%%%%%%%%%%%%%%%%%%%%%%%%%%%%%%%%%%%%

% Don't change these lines
\bsp	% typesetting comment
\label{lastpage}
\end{document}